\documentclass[12pt]{article}
\usepackage{graphicx,wrapfig}
\usepackage{amsfonts}
\usepackage{amsmath}
\usepackage{amssymb}

\def\bec{\begin{center}}
\def\eec{\end{center}}
\def\beq{\begin{equation}}
\def\eeq{\end{equation}}
\def\benu{\begin{enumerate}}
\def\eenu{\end{enumerate}}
\def\bit{\begin{itemize}}
\def\eit{\end{itemize}}
\def\beqr{\begin{eqnarray}}
\def\eeqr{\end{eqnarray}}
\def\beqrs{\begin{eqnarray*}}
\def\eeqrs{\end{eqnarray*}}
\def\btab{\begin{tabbing}}
\def\etab{\end{tabbing}}
\def\btable{\begin{tabular}}
\def\etable{\end{tabular}}
\def\rarw{\rightarrow}
\def\Rarw{\Rightarrow}
\def\om{\omega}
\def\gm{\gamma}

\def\lm{\lambda}

\def\dl{\delta}
\def\Dl{\Delta}
\def\sg{\sigma}
\def\Om{\Omega}

\def\rarw{\rightarrow}

\def\del{\partial}
\def\alp{\alpha}
\def\bt{\beta}

\def\half{\frac{1}{2}}

\def\noi{\noindent}

 \fontfamily{ptm}\selectfont
 \usepackage{mathptmx}           

\title{\bf Optical Diffraction Radiation from a beam off a circular target}

\author{Tanaji Sen \\ Accelerator Physics Center, FNAL \\ PO Box 500 \\
Batavia, IL 60510}

\date{}

\begin{document}

\maketitle

\tableofcontents

\begin{abstract}
We calculate the optical diffraction radiation generated by a bunch of
high energy particles as they pass through a round hole within an
annular metallic ring. We derive expressions for the differential angular
spectrum in the far-field and the intensities of the horizontal and 
vertical polarizations.
The sensitivity of the spectrum to changes in beam size and position is
shown. The total photon yield from the bunch is calculated and used to set
limits on the detectable wavelengths. 
\end{abstract}

\section{Introduction}

The use of optical diffraction radiation (ODR) as a diagnostic tool has
increased in recent years. The potential of this 
technique has been demonstrated in several experiments at 
KEK \cite{KEK_PRL04}, APS \cite{APS_lumpkin}, FLASH \cite{FLASH} and 
possibly other facilities. These 
experiments were performed in extraction beam lines of lepton machines.
However this technique can also
be applied to high energy hadron beams. In this report we consider the
ODR produced by such beams with the target as a round hole and apply the
results to the Tevatron.

This radiation is produced when a beam passes in the vicinity of a 
conducting target. The electro-magnetic fields due to the beam induce
currents on the target and as the beam propagates, the currents change
in time producing radiation both in the direction of beam propagation
and along the direction of specular reflection from the target. This
latter radiation, also termed backward diffraction radiation (BDR),
is more useful for diagnostics since it can be directed out at the
same longitudinal location as the target. This radiation is different
from optical transition radiation (OTR) in which the beam passes through
a metal target. Transition radiation is not suitable for continuous
monitoring of a beam in a collider due to the beam energy loss and 
emittance growth and the fact that the target may be damaged. However
the techniques for analyzing ODR are similar in many respects to
those for OTR. 

Measurements of the radiation intensity either in the near field or
far-field have been used to determine beam positions and sizes. 
For example, the beam size and beam position of a 1.28 GeV electron beam
were measured in an extraction beam line at KEK \cite{KEK_PRL04} using 
the far-field angular distribution of the radiation. The near-field 
image was used to monitor the relative
beam size of a 7 GeV electron beam in the extraction
line at APS \cite{APS_lumpkin}. In principle, 
measurements of the beam divergence are also possible using the
interference of ODR between two targets, as has been done with OTR.

This paper is motivated by the desire to use this technique in
colliders, especially for the LHC and possibly for future colliders
envisaged such as the muon collider. A brief report on these prospects 
was presented earlier \cite{Sen_PAC07}. 
If the technique yields beam measurements with sufficient accuracy and
reliability then the non-invasive nature would allow continuous
 monitoring during the length of a luminosity run. This would be 
valuable if the beam can be imaged close to the interaction points.

Synchrotron radiation is already used as a non-invasive diagnostic 
tool in the Tevatron and will also be used in the LHC. The principal
advantage of ODR is that it can be generated in a straight section
and therefore used for imaging in an experimental insertion. The 
disadvantage is that the ODR flux is less copious than synchrotron
radiation (OSR) and imaging will take longer than with OSR.

In Section 2 we briefly discuss the parameters of different hadron
colliders. In Section 3 we derive the basic results for the angular
differential spectrum of ODR from a round hole due to a bunch. We
apply these results in Section 4 to find the sensitivity of the spectrum
to beam size and offset changes. In Section 5, we calculate the expected
photon yield from a bunch per turn as a function of frequency and we
use this to find the frequency range where a sufficiently strong ODR
signal can be obtained. In Section 6 we do a brief comparison of the
ODR spectrum with the OTR spectrum. 
We briefly list in Section 7 the experimental
issues associated with measuring ODR when two beams are present. We end
with our conclusions in Section 8. We will use CGS units throughout.

\section{Hadron colliders}

Optical transition radiation (OTR) has been used in the Tevatron at 
injection energy to image the beam \cite{Scarpine}. At collision this 
technique is not feasible both because of the impact on beam quality via
multiple scattering in the target and the damage to the target itself.
However ODR is non-intercepting and has the potential to be a useful
diagnostic tool at collision energy. This technique also has potential
in the LHC where we envision that placing ODR targets on both sides of 
the interaction point
(IP) and before the first interaction region quadrupole would allow a 
non-invasive measurement of the beam size at the IP. At RHIC the 
energy is lower so one would have to use longer wavelength ODR for 
a substantial radiation flux. 

Table \ref{table: hadroncoll} shows some of the key parameters for these
hadron colliders. The beam size in the Tevatron was calculated at C0 while
for RHIC and the LHC, locations in front of the first interaction
region quadrupole were chosen. The ratio of the beam divergence to the 
opening angle of the radiation ($\sim 1/\gm$) is very small in all the 
colliders, hence the distortion 
of the spectrum due to the beam divergence should be negligible.

\begin{table}
\bec
\btable{|c|c|c|c|}  \hline
 & Tevatron & RHIC & LHC \\ \hline
Energy [GeV]  &  980 &  250 & 7000 \\
Bunch intensity & 2.7$\times 10^{11}$ & 2$\times 10^{11}$ & 
1.1$\times 10^{11}$  \\
Beam size [$\mu$m] & 400 & 1012 & 807  \\
Beam div/opening angle & 2.9$\times 10^{-3}$ & 1.2$\times 10^{-5}$ & 
5.7$\times 10^{-3}$ \\
Number of bunches & 36 & 55 $\rarw$ 120 & 2808 \\
Revolution frequency [kHz] & 47.6 & 78.2 & 11.2 \\
\hline
\etable
\eec
\caption{Table of parameters for hadron colliders}
\label{table: hadroncoll}
\end{table}

\section{ODR from a round hole}
\label{sec: round}

The fields induced by a beam as it passes through a hole depends on
the beam energy, the beam size, the beam position relative to the center
of the hole and the shape of the hole. In this paper we consider a round
hole as a target. First we analyse the fields from a single particle
and generalize results obtained many years ago by Ter-Mikaelian 
\cite{Mikaelian}. Next we consider the ODR fields and spectrum generated
by a Gaussian bunch of particles. 

\subsection{Single particle fields}
\label{subsec: singlepart}

Consider the case where a single particle moving at constant velocity $v$
goes through a round annulus made of conducting material with inner and
outer radii of $a_{in}$ and $a_{out}$ respectively. The fields of the particle
induce fields on the surface of the annulus. 
We introduce the Fourier transform of the fields as
\beq
E_x = \int E_{\om,x} e^{-i \om t} d\om \;\;\;\; 
E_y = \int E_{\om,y} e^{-i \om t} d\om
\eeq

The Fourier transformed transverse fields of a particle moving at constant velocity
along the $z$ axis are given by \cite{Mikaelian}
\beqr
 E_{\om,x} = \frac{q\alp}{\pi v} \frac{x}{\rho} e^{i\om z/v} 
K_1(\alp \rho) \nonumber \\
 E_{\om,y} = \frac{q\alp}{\pi v} \frac{y}{\rho} e^{i\om z/v} 
K_1(\alp \rho) 
\eeqr

The origin of coordinates is at the center of the hole and in particular
the hole is in the $z=0$ plane. $q$ is the particle charge, and
\[ \alp = \frac{\om}{v \gm} = \frac{k}{\gm}, \;\;\; \rho = (x^2 + y^2)^{1/2} \]
$K_1$ is a modified Bessel function of order one. 

\subsubsection{Particle at the center of target}
\label{subsubsec: 0off}

We will first consider the simpler case of the particle moving the center
of the target. We will derive equations for the fields and angular
spectral distribution which will serve as a useful check of the more
general case when the particle is offset from the center of the hole.

Consider the field at any arbitrary point on the surface of the hole.
Given the axial symmetry of the target, we use polar coordinates. The
coordinates of a point $(x,y)$ on the target are
\[ x = \rho \cos\phi, \;\;\; y = \rho \sin\phi \]
Then
\beq
\left[ \begin{array}{c} E_{\om,x} \\ E_{\om,y} \end{array} \right]
 = \frac{q\alp}{\pi v} K_1(\alp \rho)
\left[ \begin{array}{c} \cos\phi \\ \sin\phi \end{array} \right]
\eeq

We will calculate the fields at an arbitrary location using scalar
diffraction theory. Within this approximation of assuming scalar 
diffraction theory to be valid, the fields from the entire target at
an arbitrary observation point P can be found by integrating over the
annulus
\beqr
\left[ \begin{array}{c} E_{\om,x} \\ E_{\om,y} \end{array} \right]
 = -\frac{i k}{2\pi} \frac{q\alp}{\pi v}
\int_{a_{in}}^{a_{out}} \rho d\rho \int_0^{2\pi} d\phi
\frac{e^{i k R'}}{R'} K_1(\alp \rho)
\left[ \begin{array}{c} \cos\phi \\ \sin\phi \end{array} \right]
\eeqr
where $R'$ is the distance from the point on the target to the observation
point P. If $(x,y,z)$ are the coordinates of P, then
\beq
 R' = [ (x - \rho\cos\phi)^2 + (y - \rho\sin\phi)^2 + z^2]^{1/2}
\eeq

\vspace{2em}

\noi \underline{Far field spectrum}

The observation point P is assumed to be sufficiently far from the
target so that all points on the target have nearly the same phase from 
P. In this case 
the linear dimensions of the target are small compared to the distance
from the target.  This is the regime of Fraunhoffer diffraction.

If $R$ is the distance from the center of the hole 
to the point P, i.e. $R = [x^2 + y^2 + z^2]^{1/2}$, then we assume here 
that $a_{in}, a_{out} \ll R$. Thus in the phase term $e^{ikR'}$ we 
expand
\[ R' \simeq R[ 1- \frac{2\rho}{R^2}(x\cos\phi+y\sin\phi)]^{1/2}
 \simeq R - \rho \sin\theta_P \cos(\phi_P - \phi) 
\]
where we define
\beq
x = \rho_P\cos\phi_P, \;\;\; y =\rho_P\sin\phi_P, \;\;\; 
sin\theta_P = \frac{\rho_P}{R}, \;\;\; \bar{k} = k \sin\theta_P
\eeq

Then
\beqr
\left[ \begin{array}{c} E_{\om,x} \\ E_{\om,y} \end{array} \right]
 = -\frac{i k}{2\pi} \frac{q\alp}{\pi v} \frac{e^{i k R}}{R}
\int_{a_{in}}^{a_{out}} \rho d\rho \int_0^{2\pi} d\phi
 K_1(\alp \rho) \exp[- i \bar{k} \rho\cos(\phi_P - \phi)]
\left[ \begin{array}{c} \cos\phi \\ \sin\phi \end{array} \right]
\eeqr

To do the $\phi$ integrals, we use the integral representation of 
the integer Bessel functions
\beq
J_n(z) = \frac{i^n}{2\pi}\int_0^{2\pi} \exp[-i z \cos\phi] \cos n\phi d\phi
\eeq
Then
\beq
\int_0^{2\pi} \exp[- i \bar{k} \rho\cos(\phi_P - \phi)]
\left[ \begin{array}{c} \cos\phi \\ \sin\phi \end{array} \right]
 = \frac{2\pi}{i}J_1(\bar{k}\rho)
\left[ \begin{array}{c} \cos\phi_P \\ \sin\phi_P \end{array} \right]
\eeq

The integral over the radius yields
\beqr
\int_{a_{in}}^{a_{out}} \rho d\rho K_1(\alp \rho) J_1(\bar{k}\rho) & = & 
\frac{1}{\bar{k}^2+\alp^2}
\left\{ a_{out}[\bar{k}J_2(\bar{k}a_{out})K_1(\alp a_{out}) - \alp J_1(\bar{k}a_{out})K_2(\alp a_{out}) ] \right. \nonumber \\
& & \left.  - a_{in} [\bar{k}J_2(\bar{k}a_{in})K_1(\alp a_{in}) - 
\alp J_1(\bar{k}a_{in})K_2(\alp a_{in}) ]
\right\}
\eeqr
Using the recurrence relations
\[
x J_{n+1}(x) = 2n J_n(x) - xJ_{n-1}(x), \;\;\; x K_{n+1}(x) = 2n K_n(x) - xK_{n-1}(x)
\]
we can write
\beq
\int_{a_{in}}^{a_{out}} \rho d\rho K_1(\alp \rho) J_1(\bar{k}\rho) = \frac{1}{\bar{k}^2+\alp^2}
\left[ T(a_{out}; \bar{k}) - T(a_{in}; \bar{k}) \right]
\eeq
where
\beq
 T(a; \bar{k}) = -a [\bar{k}J_0(\bar{k}a)K_1(\alp a) + \alp J_1(\bar{k}a)K_0(\alp a)]  \label{eq: T_0off}
\eeq

Thus the fields are
\beq
\left[ \begin{array}{c} E_{\om,x} \\ E_{\om,y} \end{array} \right]
 = - \frac{k q\alp}{\pi v} \frac{e^{i k R}}{R}
\frac{1}{\bar{k}^2+\alp^2}\left[T(a_{out}; \bar{k}) - T(a_{in}; \bar{k}) \right]
\left[ \begin{array}{c} \cos\phi_P \\ \sin\phi_P \end{array} \right]
\label{eq: E_om_zerooff_far}
\eeq

The Poynting vector is
\beq
\vec{S}  =  \frac{c}{4\pi} \vec{E} \times \vec{B}^* 
=  \frac{c}{4\pi}[ -\bt E_z E_x^* \hat{x} - \bt E_z E_y^* \hat{y} + 
\bt(|E_x|^2 + |E_y|^2)\hat{z} ]
\eeq
where we have used 
\[ B_x = -\bt E_y, \;\;\; B_y = \bt E_x, \;\;\; B_z = 0 \]

The total energy deposited by the fields onto an element of area $dA$ is
the time integral of the projecteg Poynting vector
\beq
\frac{dW}{dA} = \int_{-\infty}^{\infty} dt \; \vec{S} \cdot \hat{n}
\eeq
where $\hat{n}$ is the unit normal to the element. For an element orthogonal to
the direction of propagation or direction of specular reflection
\beq
\frac{dW}{dA} = \int_{-\infty}^{\infty} dt S_z = \frac{\bt c}{4\pi}
\int_{-\infty}^{\infty} dt (|E_x|^2 + |E_y|^2) = 2\pi \int d\om 
[|E_{\om,x}|^2+ |E_{\om,y}|^2]
\eeq
Hence the differential angular spectrum is
\beq
\frac{d^2 W}{d\Om d\om} = \half \bt c R^2 [|E_{\om,x}|^2+ |E_{\om,y}|^2]
\eeq
where $d\Om$ is the solid angle subtended by the element at a distance $R$ from the source.

Thus in the far field, the differential spectrum is
\beq
\frac{d^2 W}{d\Om d\om} = \half \bt c (\frac{k q \alp}{\pi v})^2 
\frac{1}{[\bar{k}^2+\alp^2]^2}
\left[T(a_{out}; \bar{k}) - T(a_{in}; \bar{k}) \right]^2
\label{eq: angspect_1part_0off}
\eeq

Define a critical frequency $\om_c$, and dimensionless parameters $u, t, g$ as
\beq
\om_c = \frac{\gm c}{a_{in}}, \;\;\; u = \frac{\om}{\om_c}, \;\;\; t = \gm \sin\theta_p, \;\;\;
g = \frac{a_{out}}{a_{in}}  \label{eq: scaledvar}
\eeq
Then other parameters can be written in terms of these dimensionless 
parameters as $ k = \gm u/(\bt a_{in})$, $\alp= u/(\bt a_{in})$ etc and
\beqrs
\left[T(a_{out}; \bar{k}) - T(a_{in}; \bar{k}) \right]^2 
 & = & (\frac{u}{\bt})^2 \left\{
g[ t J_0(\frac{1}{\bt} gut) K_1(\frac{1}{\bt} gu) + J_1(\frac{1}{\bt}gut)
 K_0(\frac{1}{\bt} gu)] \right. \\
& & \left. - 
[ t J_0(\frac{1}{\bt} ut) K_1(\frac{1}{\bt} u) + J_1(\frac{1}{\bt}ut)
 K_0(\frac{1}{\bt} u)] \right\}
\eeqrs

The angular spectral distribution thus is
\beqr
\frac{d^2 W}{d\Om d\om} & = & 
\frac{\bt c}{2}(\frac{q\gm}{\pi v})^2 
\frac{u^2}{[1+t^2]^2}
 \left\{
g[ t J_0(\frac{1}{\bt} gut) K_1(\frac{1}{\bt} gu) + J_1(\frac{1}{\bt}gut)
 K_0(\frac{1}{\bt} gu)] \right. \nonumber \\ 
& & \left. - 
[ t J_0(\frac{1}{\bt} ut) K_1(\frac{1}{\bt} u) + J_1(\frac{1}{\bt}ut)
 K_0(\frac{1}{\bt} u)] \right\}^2  \label{eq:d2W_far_0offset}
\eeqr
We comment on some features of this expression
\bit
\item This spectrum depends on the magnitude of the 
inner radius $a_{in}$ only through the critical frequency $\om_c$.
\item The main dependence of the spectrum on the size of the target
is through the dimensionless ratio $g = a_{out}/a_{in}$. This is important
since it suggests that the target hole may be enlarged to allow more
space for the beam while at the same time increasing the outer radius
without changing the spectrum. The parameter that will change in this
case is the critical frequency $\om_c$ and consequently the dimensional
frequency $\om$. 
\eit

Th differential spectrum may be found by integrating over the solid angle
\beq
\frac{dW}{d\om} = \int \frac{d^2 W}{d\Om d\om} d\phi sin\theta_P d\theta_P
 = \frac{4\pi}{\gm^2} \int_0^{\gm} 
\frac{d^2 W}{d\Om d\om} \frac{t}{\sqrt{1 - t^2/\gm^2}} dt
\eeq

Define the function
\beqr
F(g,u) & = & \int_{0}^{\gm} dt \frac{t}{\sqrt{1 - t^2/\gm^2}}
\frac{1}{[1+t^2]^2} \left\{
g[ t J_0(\frac{1}{\bt} gut) K_1(\frac{1}{\bt} gu) + J_1(\frac{1}{\bt}gut)
 K_0(\frac{1}{\bt} gu)] \right. \nonumber \\
& & \left. - 
[ t J_0(\frac{1}{\bt} ut) K_1(\frac{1}{\bt} u) + J_1(\frac{1}{\bt}ut)
 K_0(\frac{1}{\bt} u)] \right\}^2
\eeqr

The number of photons $\Dl N$ emitted {\em by a single charged particle}
into a bandwidth $\Dl \om$ is
\beq
\Dl N \equiv \frac{d N}{d\om} \Dl \om =
 (\frac{1}{\hbar \om}\frac{d W}{d\om}) \Dl \om
\eeq
which can be written as
\beq
\fbox{\rule[-.5cm]{0cm}{1cm} 
$\displaystyle 
\Dl N_{ph} = (\frac{1}{\pi \bt} \alp_f) u^2 F(g,u) \frac{\Dl \om}{\om} $}
\label{eq: round_single_far}
\eeq
where $\alp_f = q^2/(\hbar c) \approx 1/137$ is the fine structure
constant. This depends on the relative frequency $u = \om/\om_c$ and
the relative bandwidth $\Dl\om/\om$.

\vspace{2em}

\noi \underline{\sf Example: Tevatron parameters}

\begin{figure}
\centering
\includegraphics{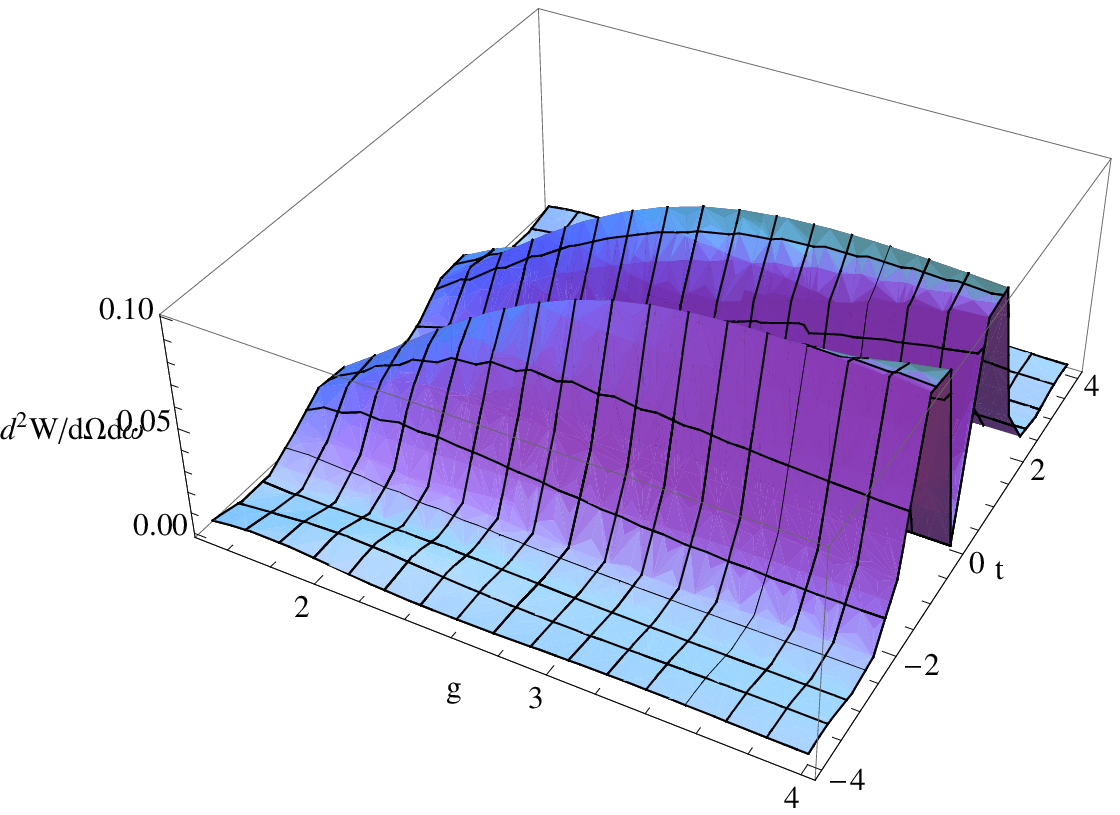}
\caption{The differential angular spectrum as a function of the ratio
of the outer and inner radii of the target $g = a_{out}/a_{in}$ and the angular
variable $t=\gm\sin\theta_P$ at constant $u=1$. Note that the spectrum 
saturates as a function of $g$ for $g > 3$.}
\label{fig: Fig_Wgt3D}
\vspace{1em}
\includegraphics{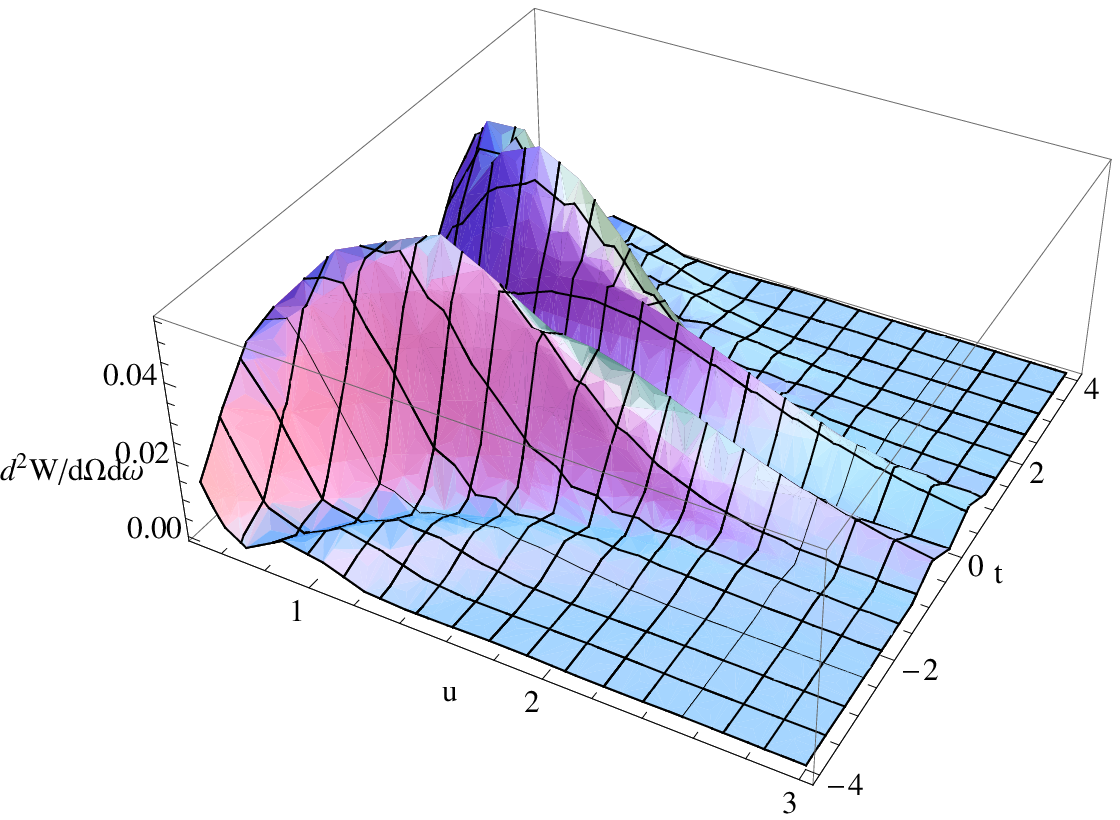}
\caption{The differential angular spectrum as a function of the ratio
$u = \om/\om_c$ and the angular variable $t=\gm\sin\theta_P$ at constant
$g=1.5$. The spectrum peaks close to $u=1$, i.e. close to the critical 
frequency $\om_c$.}
\label{fig: Fig_Wut3D}
\end{figure}

We evaluate the spectrum and the number of photons per particle for the 
Tevatron. Energy = 980 GeV, number of particles per bunch 
$N_p = 2.7\times 10^{11}$. Figure \ref{fig: Fig_Wgt3D} shows the
differential angular spectrum as a function of the parameters $g,t$ 
with the frequency fixed at $\om=\om_c$. Figure \ref{fig: Fig_Wut3D}
shows the spectrum as a function of the parameters $u, t$ at a fixed
ratio $g=1.5$. See the figure captions for comments.

\begin{figure}
\centering
\includegraphics[width=10cm,height=6cm]{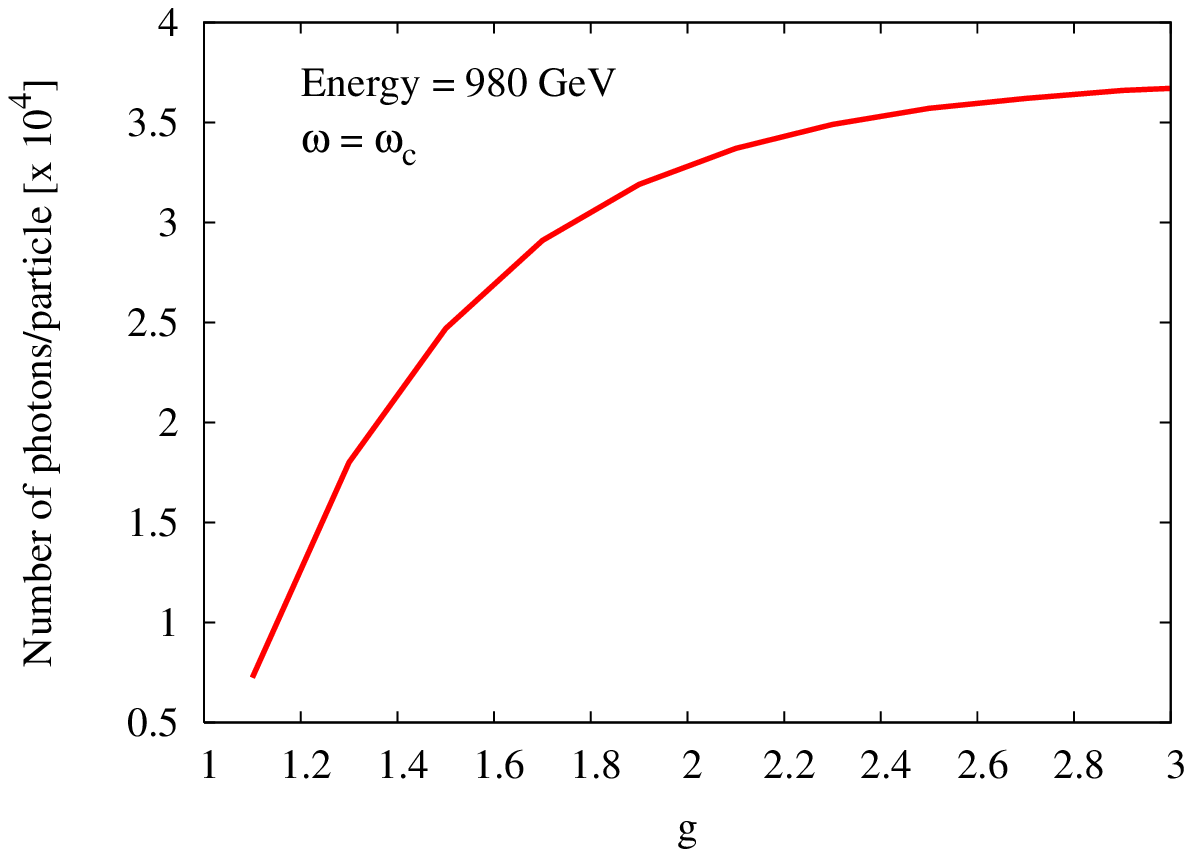}
\caption{The number of photons per particle as a function of the parameter
$g$ emitted at the critical frequency $\om_c$. We observe that for 
$g > 2$, the number of photons increases slowly with $g$.}
\label{fig: Nphot_1}
\end{figure}
Figure \ref{fig: Nphot_1} shows the number of photons per particle
calculated using the expression Eq (\ref{eq: round_single_far}) and
assuming $\om=\om_c$ and a 1\% bandwidth or $\Dl\om/\om = 0.01$. 
The curve again shows that there is little gain in intensity when
the target size increases beyond $g > 2.5$. After some 
transverse distance from the particle, its field
has dropped to sufficiently low values that no radiators
in the target can be excited and therefore there is no further increase
in the ODR radiation with increasing material in the target. 
The number of photons emitted by a single particle in one pass through
the center of the target can be found from this curve which is calculated
for the critical frequency $\om_c$. For example at 
\[ 
g = 1.1 \;\;\; \Rarw \Dl N = 1.44\times 10^{-4} 
\]
A very simple estimate for the number of photons
emitted by a bunch in a single pass at the frequency $\om_c$ is therefore
\[ \Dl N ({\rm bunch}) = N_p \times \Dl N = 3.2 \times 10^5 \]

This in fact is an underestimate since it assumes that all particles
are at the center of the target and therefore furthest from the 
material of the target. A more precise estimate using the density 
distribution of the bunch will be obtained in the following section.


\vspace{3em}

\noi{\underline{Near field spectrum}}

Here we calculate the field distribution at a distance close enough to
the target that the phase differences between different points on the
target to the observation point is significant. This is the region
of Fresnel diffraction.

Here in the expansion for $R'$ we keep the next order term in $\rho/R$.
Thus
\[ R' = R[1 - 2\frac{\rho\rho_P}{R62}\cos(\phi-\phi_P) + 
\frac{\rho^2}{R^2}]^{1/2} \simeq R - \rho\sin\theta_P\cos(\phi-\phi_P)
 + \half \frac{\rho^2}{R}
\]
The integration for the fields contains the extra phase factor
$\exp[i k \rho^2/(2R)]$ when compared to the fields calculated in the
far field approximation. 

Define the dimensionless variables
\beq
p = \frac{\rho}{a_{in}}, \;\;\; r = \frac{R}{a_{in}}, \;\;\; 
\bar{a}=\frac{a}{a_{in}}, \;\;\; g=\frac{a_{out}}{a_{in}}, \;\;\; 
\eta = \frac{\gm u}{2\bt r}, \;\;\;
\Rarw \frac{k\rho^2}{2R} = \eta p^2
\eeq

Define the complex function
\beqr
S[\bar{a}; \bar{k}, r] & = & \frac{1}{a_{in}^2}\int_0^a \rho J_1(\bar{k}\rho) K_1(\alp a)
 \exp[i \frac{k\rho^2}{2R}] d\rho \nonumber \\
& = & \int_0^{\bar{a}} p J_1(\frac{1}{\bt}ut p)K_1(\frac{1}{\bt}u p)
 \exp[i \eta p^2] dp
\eeqr

Then following similar steps as in the previous section, it follows
that the Fourier transforms of the transverse electric fields are
\beq
\left[ \begin{array}{c} E_{\om,x} \\ E_{\om,y} \end{array} \right]
 = - \frac{k q\alp}{\pi v} \frac{e^{i k R}}{R}
\left[S(g; \bar{k}, r) - S(1; \bar{k}, r) \right]
\left[ \begin{array}{c} \cos\phi_P \\ \sin\phi_P \end{array} \right]
\eeq

The angular spectral distribution is
\beq
\frac{d^2 W}{d\Om d\om} = \half \frac{q^2}{\pi^2 \bt c}(\gm u^2)^2
| \left[S(g; \bar{k}) - S(1; \bar{k}) \right] |^2
\eeq

The frequency spectrum is found by integrating over the solid angle
\beq
\frac{dW}{d\om} = \frac{4\pi}{\gm^2}\int_{0}^{\gm} 
\frac{d^2 W}{d\Om d\om} \frac{t}{\sqrt{1 - t^2/\gm^2}} dt
\eeq
Define the function
\beq
D(g,u, r) = \int_{0}^{\gm} dt \frac{t}{\sqrt{1 - t^2/\gm^2}}
| \left[S(g; \bar{k}, r) - S(1; \bar{k}, r) \right] |^2
\eeq

The number of photons emitted {\em by a single charged particle} 
into a bandwidth $\Dl\om$ is therefore
\beq
\fbox{\rule[-.5cm]{0cm}{1cm} 
$\displaystyle 
\Dl N_{ph} = (\frac{1}{\pi\bt} \alp_f) u^4 D(g,u,r) \frac{\Dl \om}{\om} $}
\eeq
This depends on the inner radius $a_{in}$ through the scaled
variables $g = a_{out}/a_{in}, r = R/a_{in}, u = \om /\om_c$.


\subsubsection{Particle offset from the center of target}
\label{subsubsec: off}

Our final aim is to find the spectral distribution from a bunch of
particles. Towards that end we first need to know the field distribution
from a particle offset from the center of the target.
\begin{figure}
\centering
\includegraphics{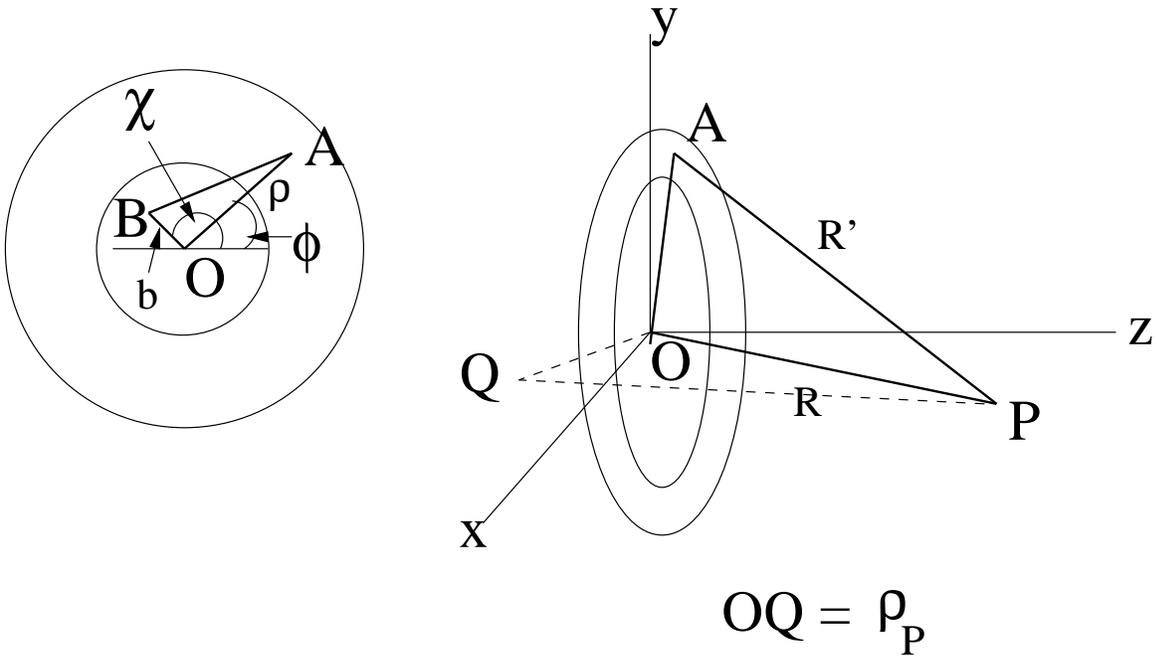}
\caption{The target centered at O is an annulus of inner radius $a_{in}$ and outer
radius $a_{out}$. B is the location of the particle offset by a distance $b$
from the center of the target, A is the arbitrary location on the 
target for the field calculation. In the figure on the right, P is the
point of
observation, Q is the projection of P on to $x-y$ plane.}
\label{fig: round}
\end{figure}
\begin{figure}
\centering
\includegraphics[width=8cm,height=5cm]{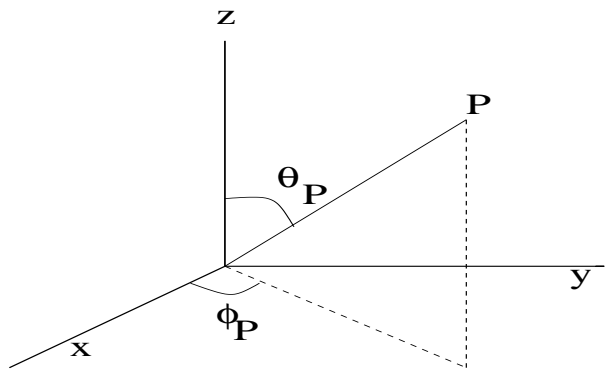}
\caption{P is the observation point, angle $\theta_P$ is the angle
with the z axis and $\phi_P$ is the angle with the x axis made by the
projection onto the x-y plane.}
\label{fig: angles}
\end{figure}
The target is in the $z=0$ plane and the particle moves with uniform
velocity $v$ along the z axis. 
Figure \ref{fig: round} shows a sketch of the target with center at O, 
the particle is at B
and the field on the target is calculated at point A. Again we use polar
coordinates: particle at B has coordinates $(b,\chi)$ while point A
has coordinates $(\rho,\phi)$ with respect to the center O. The distance
of A relative to the particle at B is 
\beq
 r_{\perp} = [\rho^2 + b^2 - 2\rho b \cos(\chi-\phi)]^{1/2}
\eeq
while the separation along the $(x,y)$ axes are individually
\beq
x = x_A - x_B = \rho\cos\phi - b\cos\chi, \;\;\;
y = y_A - y_B = \rho\sin\phi - b\sin\chi 
\eeq
Hence the Fourier transforms of the transverse fields at A are
\beq
\left[ \begin{array}{c} E_{\om,x} \\ E_{\om,y} \end{array} \right]
 = \frac{q\alp}{\pi v} \frac{K_1(\alp \rho_{\perp})}{ \rho_{\perp}}
\left[ \begin{array}{c} \rho\cos\phi - b\cos\chi \\ 
\rho\sin\phi - b\sin\chi \end{array} \right]
\eeq
As before $R'$ is the distance of the observation point P from the
location of the field, thus
\[ R' = [(x_P - \rho\cos\phi)^2 + (y_P - \rho\sin\phi)^2 + z_P^2]^{1/2}
\]
where $(x_P,y_P,z_P) = (\rho_P\cos\phi_P,\rho_P\sin\phi_P,z_P)$ are the 
coordinates of the point P. Figure \ref{fig: angles} shows the 
relevant angles $\theta_P, \phi_P$. 

Integrating over the annulus, the fields from the entire target at the
point P are (using scalar diffraction theory)
\beq
\left[ \begin{array}{c} E_{\om,x} \\ E_{\om,y} \end{array} \right]
 = -\frac{ik}{2\pi}\frac{q\alp}{\pi v} \int_{a_{in}}^{a_{out}} \int_0^{2\pi}
 \rho d\rho d\phi \frac{e^{i k R'}}{R'}
\frac{K_1(\alp \rho_{\perp})}{ \rho_{\perp}}
\left[ \begin{array}{c} \rho\cos\phi - b\cos\chi \\ 
\rho\sin\phi - b\sin\chi \end{array} \right]
\eeq


\underline{Far field spectrum}

We assume that the point P is sufficiently far from the target that
the far field approximation is valid. The phase term is expanded as
\[
\frac{e^{i kR'}}{R'} = \frac{e^{i k R}}{R}
e^{-i\bar{k}\rho\cos(\phi-\phi_P)}
\]
where $\bar{k}=k\sin\theta_P$, $\theta_P$ is the angle made by OP with the $z$ axis or $\sin\theta_P = \rho_P/R$. 

The integrations are simplified if we write the integrands
as derivatives with respect to variables that are not integrated. We
note first that
\[
\left[ \begin{array}{c}
\frac {\del}{\del b} \\ \frac {\del}{\del \chi} \end{array} \right]
K_0(\alp \rho_{\perp}) = \frac{\alp K_1(\alp \rho_{\perp})}
{\rho_{\perp}} 
\left[ \begin{array}{c}
\rho \cos(\phi - \chi) - b \\ b \rho \sin(\phi - \chi)
\end{array} \right]
\]
Next we transform the fields to a frame rotated by angle $\chi$ with
respect to the $(x,y)$ axes, i.e.
\beq
\left[ \begin{array}{c}
E_1 \\ E_2 \end{array} \right] = 
\left[ \begin{array}{cc}
\cos\chi & \sin\chi \\ -\sin\chi & \cos\chi  \end{array} \right]
\left[ \begin{array}{c}
E_{\om,x} \\ E_{\om,y}  \end{array} \right]
\eeq
Then using the derivative expressions above, we find
\beq
\left[ \begin{array}{c}
E_1 \\ E_2 \end{array} \right] = 
-\frac{ik}{2\pi}\frac{q}{\pi v}\frac{e^{i k R}}{R}
\left[ \begin{array}{c}
\frac {\del}{\del b} \\ 
\frac {1}{b} \frac {\del}{\del \chi} \end{array} \right]
 \int_{a_{in}}^{a_{out}} \int_0^{2\pi} d\rho d\phi\; \rho 
K_0(\alp\rho_{\perp}) e^{-i\bar{k}\rho\cos(\phi-\phi_P)}
\eeq
The double integral can be factorized into the product of single
integrals by using the expansions
\beqr
K_0(\alp[\rho^2+b^2-2b\rho\cos(\phi-\chi)]^{1/2}) & = &
\sum_{n=-\infty}^{\infty} I_n(\alp b) K_n(\alp \rho) e^{i n(\phi-\chi)}
\nonumber \\
\exp[-i\bar{k}\rho\cos(\phi-\phi_P)] & = & 
\sum_{n=-\infty}^{\infty} (-i)^n J_n(\bar{k}\rho)e^{i n(\phi-\phi_P)}
\eeqr
The integration over $\phi$ is trivial and the double integral reduces
to
\beqrs
 H \equiv 
 \int_{a_{in}}^{a_{out}} \int_0^{2\pi} d\rho d\phi\; \rho 
K_0(\alp\rho_{\perp}) e^{-i\bar{k}\rho\cos(\phi-\phi_P)}
&  = & 2\pi \sum_{n=-\infty}^{\infty} (-i)^n I_n(\alp b) 
e^{-i n(\chi-\phi_P)} \\
& & \times \int_{a_{in}}^{a_{out}} \rho J_n(\bar{k}\rho)
K_n(\alp\rho) d\rho
\eeqrs
The integration over $\rho$ can be done symbolically using Mathematica
\cite{Math} which yields
\beqrs
 \int_{0}^{a} \rho J_n(\bar{k}\rho) K_n(\alp\rho) d\rho 
& = & \frac{a}{\bar{k}^2 + \alp^2}[
\bar{k} J_{n+1}(\bar{k}a)K_n(\alp a) - 
 \alp J_n(\bar{k}a)K_{n+1}(\alp a) ] \\
& = & -\frac{a}{\bar{k}^2 + \alp^2}[\bar{k} J_{n-1}(\bar{k}a)K_n(\alp a)
 + \alp J_n(\bar{k}a)K_{n-1}(\alp a)]
\eeqrs
where recurrence relations for the Bessel functions were used. Then
defining a function $T_n(a;\bar{k})$ similar to the one defined
in Equation (\ref{eq: T_0off}) in Section \ref{subsubsec: 0off},
\beq
T_n(a;k) = -a[\bar{k} J_{n-1}(\bar{k}a)K_n(\alp a)
 + \alp J_n(\bar{k}a)K_{n-1}(\alp a)]
\eeq
After some further simplifications we can write
\beq
H = \frac{2\pi}{\bar{k}^2 + \alp^2} \sum_{n=0}^{\infty} C_n I_n(\alp b)
[T_n(a_{out}; \bar{k}) - T_n(a_{in}; \bar{k})] \cos n(\chi-\phi_P)
\eeq
where
\[
C_n  =  1 \;\;\; {\rm for}\;\; n = 0;   \;\;\; \;\;\; C_n=  2 (-i)^n \;\;\;
{\rm for}\;\;  n \ge 1
\]
and the rotated fields are 
\beq
\left[ \begin{array}{c}
E_1 \\ E_2 \end{array} \right] = 
-\frac{ik \alp q}{\pi v(\bar{k}^2 + \alp^2)}\frac{e^{i k R}}{R}
\sum_{n=0} C_n [T_n(a_{out};\bar{k})-T_n(a_{in};\bar{k})]
\left[ \begin{array}{c}
I_n'(\alp b) \cos[n(\chi-\phi_P)] \\ 
- n \frac{I_n(\alp b)}{\alp b} \sin[n(\chi-\phi_P)]
\end{array} \right] 
\eeq
The fields in the lab frame $(E_{\om,x}, E_{\om,y})$ are obtained by
applying the inverse rotation
\beq
\left[ \begin{array}{c}
E_{\om,x} \\ E_{\om,y}  \end{array} \right] =
\left[ \begin{array}{cc}
\cos\chi & -\sin\chi \\ \sin\chi & \cos\chi  \end{array} \right]
\left[ \begin{array}{c}
E_1 \\ E_2 \end{array} \right]   
\label{eq: E_lab_rot} 
\eeq

\underline{Limit of zero offset}

Before proceeding further, we first check that the expressions
derived here reduce to the expressions derived in the previous
sub-section in the limit that the offset is zero.

First we note that 
\[ \lim_{\chi \rarw 0} 
\left[ \begin{array}{c}
E_{\om,x} \\ E_{\om,y}  \end{array} \right] =
\left[ \begin{array}{c}
E_1 \\ E_2 \end{array} \right]
\]
Now using the fact that
\[ \lim_{b\rarw 0} \frac{I_n(\alp b)}{b} = \frac{\alp}{2}  \dl_{n,1} \]
Then only the $n=1$ term in the sum contributes and we have
\beq
\lim_{b, \chi \rarw 0} \left[ \begin{array}{c} E_{\om,x} \\ E_{\om,y} \end{array} \right]
 = -\frac{k q\alp}{\pi v(\bar{k}^2 + \alp^2)} \frac{e^{ikR}}{R}
 [T_1(a_{out};\bar{k}) - T_1(a_{in};\bar{k})] 
\left[ \begin{array}{c} \cos\phi_P \\ \sin\phi_P \end{array} \right]
\eeq
These expressions agree with the expressions in Equation 
(\ref{eq: E_om_zerooff_far}) derived earlier.

We return now to the case with offset. 
The differential angular spectrum is proportional to the square of the
absolute norm which is preserved under rotations,
\[ |E_{\om,x}|^2 + |E_{\om,y}|^2 = |E_1|^2 + |E_2|^2 \]
Hence the differential angular spectrum is given by
\beq
\frac{d^2 W}{d\Om d\om} = \half\bt c R^2 [|E_1|^2 + |E_2|^2]
\eeq
The spectrum from a single particle therefore is 
\beq
\frac{d^2 W}{d\Om d\om}|_{particle} =  \half\bt c
[\frac{k q\alp}{\pi v(\bar{k}^2 + \alp^2)}]^2 
\sum_{m=0}\sum_{n=0} C_m C_n^* V_{mn} 
\eeq
where we have defined 
\beqr
V_{mn} & = & U_{mn} \{ I_m'(\alp b) I_n'(\alp b)\cos m(\chi-\phi_{P})
\cos n(\chi-\phi_{P}) \\
& & + m n \frac{I_m(\alp b) I_n(\alp b)}{\alp^2 b^2}
\sin m(\chi-\phi_{P})\sin n(\chi-\phi_{P}) \} \\
U_{mn} & = & [T_m(a_{out};\bar{k}) - T_m(a_{in};\bar{k})] 
[T_n(a_{out};\bar{k}) - T_n(a_{in};\bar{k})] 
\label{eq: V_Umn}
\eeqr
It is helpful to use the scaled variables $g,u,t$ introduced in 
Equation (\ref{eq: scaledvar}). Then
\beq
k  =  \frac{\gm u}{\bt a_{in}}, \; \bar{k} = k\sin\theta_P = 
\frac{tu}{\bt a_{in}}, \;\;\; \alp = \frac{u}{\bt a_{in}}, \;\;\;
\left(\frac{k q\alp}{\pi v(\bar{k}^2 + \alp^2)}\right)^2  =  
(\frac{q \gm}{\pi v})^2 \frac{1}{(1+t^2)^2}
\eeq
Furthermore
\beqr
[T_m(a_{out};\bar{k}) - T_m(a_{in};\bar{k})] & = & -\frac{u}{\bt}
\left\{ g[tJ_{m-1}(\frac{gut}{\bt})K_m(\frac{gu}{\bt})+
         J_{m}(\frac{gut}{\bt})K_{m-1}(\frac{gu}{\bt})] \right. \nonumber \\
& &    \left. - [tJ_{m-1}(\frac{ut}{\bt})K_m(\frac{u}{\bt})+
         J_{m}(\frac{ut}{\bt})K_{m-1}(\frac{u}{\bt})] \right\}
\eeqr
Hence $U_{mn}(g,u,t)$ is only a function of the scaled variables
$g,u,t$ but does not depend on the inner radius $a_{in}$. Similarly
we define the scaled offset
\[ b_s = \frac{b}{a_{in}}, \;\;\; \Rightarrow \alp b = b_s\frac{u}{\bt} \]
This shows that $V_{mn}(g,u,t,b_s)$ is also independent of the inner 
aperture $a_{in}$.

Since $V_{mn}$ is real and symmetric under the interchange of indices, we
can write 
\[ \sum_{m=0}\sum_{n=0} C_m C_n^* V_{mn}  = \half
\sum_{m=0}\sum_{n=0} (C_m C_n^* + C_m^* C_n) V_{mn}  
\]
Using the definition of $C_m$ we obtain
\beqrs
C_m + C_m^* & = & 0 \;\;\;\;\;\;\; m\; {\rm odd}  \\
 & = & 4(-1)^{m/2}  \;\;\;\;\;\;\; m\; {\rm even} \\
C_m C_n^* + C_m^* C_n & = & 0 \;\;\;\;\;\;\; m-n\; {\rm odd}  \\
 & = & 8(-1)^{(m-n)/2} \;\;\;\;\;\;\; m-n\; {\rm even}
\eeqrs
Hence
\beqr
\frac{d^2 W}{d\Om d\om}|_{particle}(g,u,t,b,\chi) & = & \half\bt c
(\frac{q \gm}{\pi v})^2 \frac{1}{(1+t^2)^2}
[ V_{00} + 4 \sum_{m=2,4,...} (-1)^{m/2} V_{m0} \nonumber \\
& &  + 4 \sum_{m=1} \sum_{n=1; \; |m-n|=even} (-1)^{(m-n)/2} V_{mn} ]
\label{eq: spectrum_1part}
\eeqr
In the limit that the offset goes to zero, this reduces to 
\beq
\lim_{b\rarw 0}
\frac{d^2 W}{d\Om d\om}|_{particle} =  \half\bt c
(\frac{q \gm}{\pi v})^2 \frac{1}{(1+t^2)^2} U_{11}
\eeq
which agrees with the expression in Equation (\ref{eq:d2W_far_0offset}).

\begin{figure}
\centering
\includegraphics[height=6cm,width=10cm]{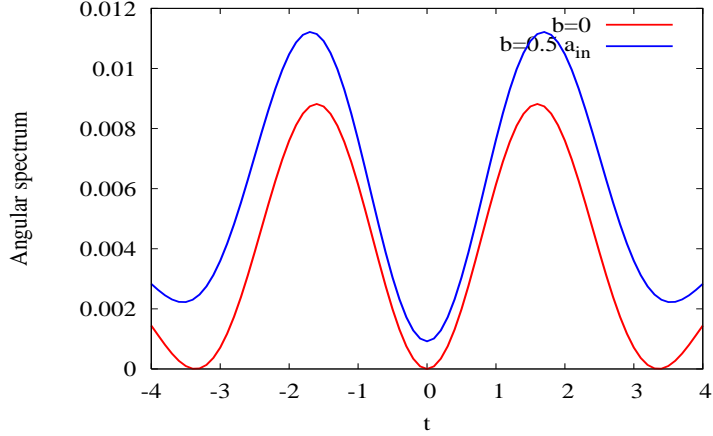}
\caption{The differential angular spectrum from a single particle vs 
the angle of observation ($t=\gm\sin\theta_P$) with
two different offsets $b=0$ and $b=0.5 a_{in}$. Other parameter values are
fixed at $g=1.3, u = 1, \chi = 0, \phi_P=\pi/4$. With non-zero offset, both the minimum and the maximum value of the flux increases. Note that
the position of th maximum also changes with the offset.}
\label{fig: W1part_b_0_0.5}
\end{figure}

The dependence of the single particle spectrum on the parameters 
$(g,u,t)$ is similar to that seen in Figures \ref{fig: Fig_Wgt3D} and 
\ref{fig: Fig_Wut3D}. Figure \ref{fig: W1part_b_0_0.5} shows the
dependence of the single particle spectrum on the angle of observation
for two different offsets $b$ from the center of the hole. With an
increased offset the particle is closer to the material of the target resulting
in a larger radiation flux . 

The intensities of the different polarizations can be found from the 
components of the electric field. From Equation \ref{eq: E_lab_rot} it
follows that
\beqr
E_{\om,x} & = & -\frac{i k \alp q}{\pi v(\bar{k}^2 + \alp^2)} \frac{e^{ikR}}{R}
\sum_n C_n [T_n(a_{out};k) - T_n(a_{in};k)] \nonumber \\
& & \times \left\{I_n'(\alp b) \cos[n(\chi-\phi_P)]\cos\chi + 
n \frac{I_n(\alp b)}{\alp b} \sin[n(\chi-\phi_P)]\sin\chi \right\} \\
E_{\om,y} & = & -\frac{i k \alp q}{\pi v(\bar{k}^2 + \alp^2)} \frac{e^{ikR}}{R}
\sum_n C_n [T_n(a_{out};k) - T_n(a_{in};k)] \nonumber \\
& & \times \left\{I_n'(\alp b) \cos[n(\chi-\phi_P)]\sin\chi -
n \frac{I_n(\alp b)}{\alp b} \sin[n(\chi-\phi_P)]\cos\chi \right\}
\eeqr
Hence it follows that the intensities of the two polarizations are given by
\beqr
\frac{d^2 W^{x,y}}{d\Om d\om}|_{particle}(g,u,t,b,\chi) & = & \half\bt c
(\frac{q \gm}{\pi v})^2 \frac{1}{(1+t^2)^2}
[ V_{00}^{x,y} + 4 \sum_{m=2,4,...} (-1)^{m/2} V_{m0}^{x,y} \nonumber \\
& &  + 4 \sum_{m=1} \sum_{n=1; \; |m-n|=even} (-1)^{(m-n)/2} V_{mn}^{x,y} ]
\label{eq: Ang_Wxy_part}
\eeqr
where 
\beqr
V_{mn}^{x} & = & U_{mn} [I_m'(\alp b) \cos[m(\chi-\phi_P)]\cos\chi + 
m \frac{I_m(\alp b)}{\alp b} \sin[m(\chi-\phi_P)]\sin\chi] \nonumber \\
& & \times [I_n'(\alp b) \cos[n(\chi-\phi_P)]\cos\chi + 
n \frac{I_n(\alp b)}{\alp b} \sin[n(\chi-\phi_P)]\sin\chi] \\
V_{mn}^{y} & = & U_{mn} 
[I_m'(\alp b) \cos[m(\chi-\phi_P)]\sin\chi -
m \frac{I_m(\alp b)}{\alp b} \sin[m(\chi-\phi_P)]\cos\chi] \nonumber \\
& & \times [I_n'(\alp b) \cos[n(\chi-\phi_P)]\sin\chi -
n \frac{I_n(\alp b)}{\alp b} \sin[n(\chi-\phi_P)]\cos\chi
\eeqr

We will use the polarized intensities to examine their sensitivity to
beam parameters in the next section.


\subsection{Spectrum from a bunch}

So far we've dealt with the spectrum from a single particle traveling
through the hole. Now we'll consider the spectrum from a typical 
bunch. We assume here a Gaussian
distribution of N particles with transverse rms sizes $\sg_x,\sg_y$. We
also assume that the bunch center is offset from the center of the
target with offsets $(x_0, y_0)$. Then
\beqr
\rho(x,y) & = & \frac{N}{2\pi\sg_x\sg_y} \exp[-\frac{(x-x_0)^2}{2\sg_x^2}-
\frac{(y-y_0)^2}{2\sg_y^2}] \nonumber \\
& = & \frac{N}{2\pi\sg_x\sg_y} \exp[-\frac{(b_s\cos\chi-x_{0,s})^2}{2\sg_{x,s}^2}-
\frac{(b_s\sin\chi-y_{0,s})^2}{2\sg_{y,s}^2}] \\
& \equiv &   \frac{N}{2\pi\sg_x\sg_y}
 \rho_s(b_s,\chi,x_{0,s},y_{0,s},\sg_{x,s},\sg_{y,s}) 
\label{eq: rho_bs_chi}
\eeqr
The last equality defines the scaled density $\rho_s$ and 
we have scaled the other variables by $a_{in}$, the inner radius of the hole,
\beq
b_s = \frac{b}{a_{in}}, \; x_{0,s} = \frac{x_0}{a_{in}}, \;
y_{0,s} = \frac{y_0}{a_{in}}, \; \sg_{x,s} = \frac{\sg_{x}}{a_{in}}, \;
\sg_{y,s} = \frac{\sg_{y}}{a_{in}}
\label{eq: scal_a_in}
\eeq

The differential angular spectrum averaged over the bunch distribution is
\beqr
\frac{d^2 W}{d\Om d\om}|_{bunch} &= &\int b db \int d\chi \;\; \rho(b,\chi)
\frac{d^2 W}{d\Om d\om}|_{particle}
 = a_{in}^2 \int b_s db_s \int d\chi \;\; \rho(b_s,\chi)
\frac{d^2 W}{d\Om d\om}|_{particle} \nonumber \\
& = & \frac{N}{2\pi\sg_{x,s}\sg_{y,s}} \int b_s db_s \int d\chi \;\; \rho_s
\frac{d^2 W}{d\Om d\om}|_{particle}
\label{eq: Ang_W_bunch}
\eeqr
It is important to note that the spectrum depends only on the scaled
variables introduced in Equation (\ref{eq: scal_a_in}) but not on the
absolute values of $a_{in}$, $(\sg_x,\sg_y)$, $(x_0, y_0)$. Hence
this is a universal expression; the only dependence on machine specific
parameters is on the beam energy and the bunch intensity. 

The two-dimensional integrals over $(b_s,\chi)$ can be factored
as the product of single integrals over $b_s$ and $\chi$ individually.
However the integration over $\chi$ introduces a triple summation and 
the integrals over $b$ cannot be performed analytically. 
Instead we will evaluate the integrations over the bunch numerically.

\begin{figure}
\centering
\includegraphics[height=6cm,width=10cm]{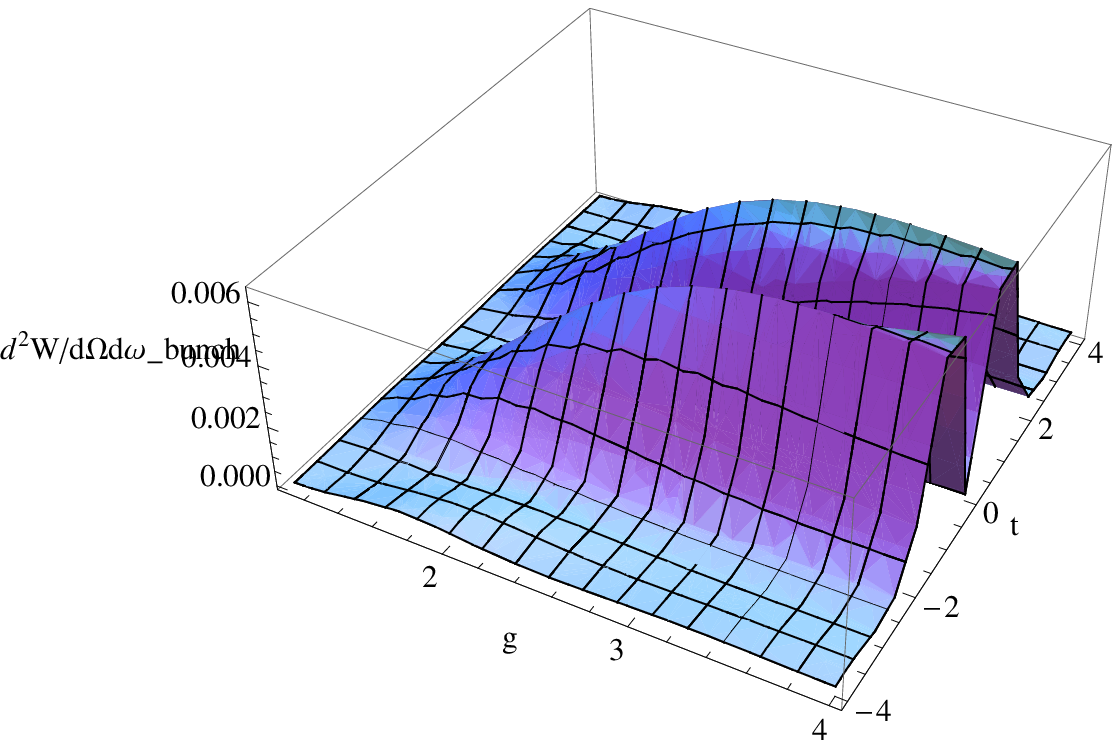}
\caption{The differential angular spectrum from a bunch as a function of 
$g = a_{out}/a_{in}$ and the angular variable $t=\gm\sin\theta_P$ at 
constant $u=1$. As a function of $g$, the spectrum is relatively flat
for $g > 2.5$.}
\label{fig: Wbunch_gt}
\vspace{1em}
\parbox[h]{9cm}{
\includegraphics[height=6cm,width=8cm]{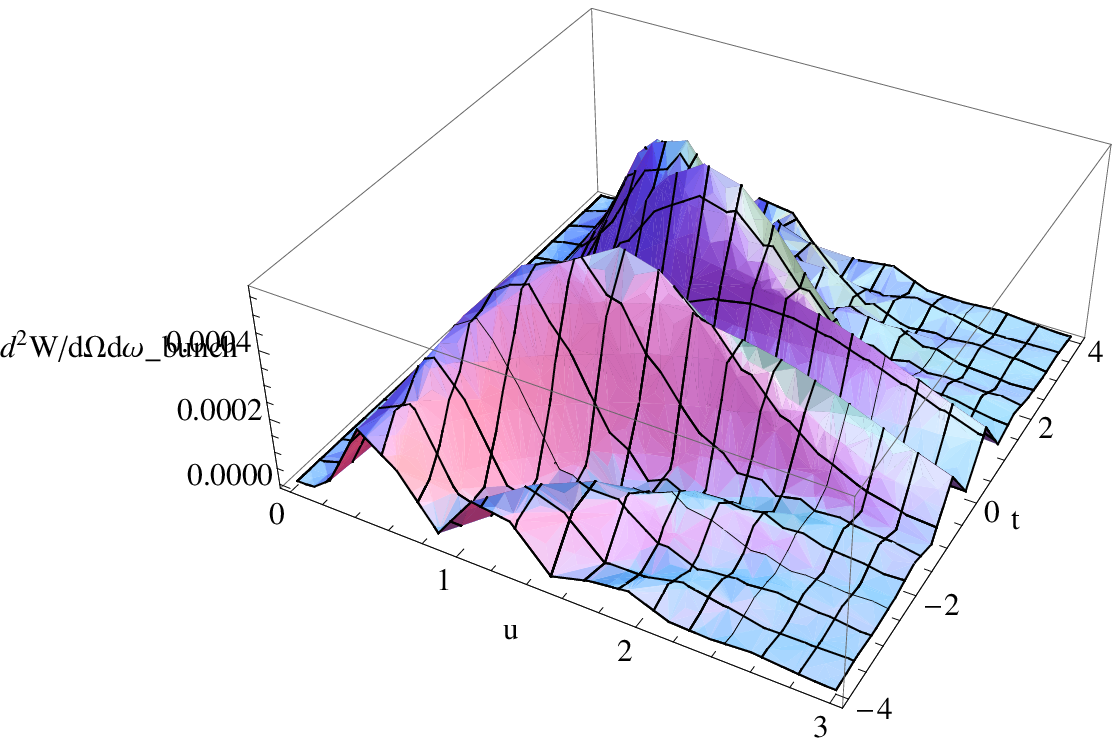}}
\parbox[h]{7cm}{
\includegraphics[height=6cm,width=8cm]{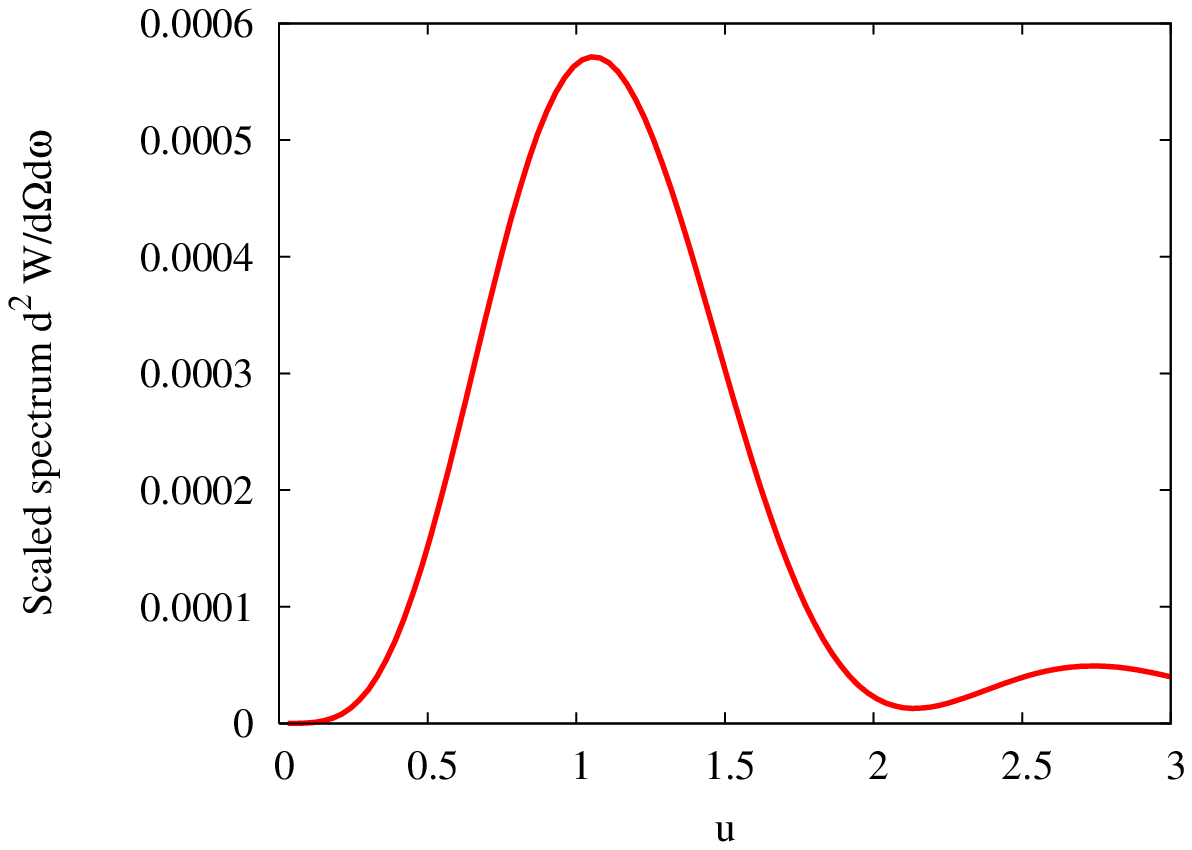}}
\caption{Top: The differential angular spectrum from a bunch as a function of 
the scaled frequency $u$ and the angular
variable $t=\gm\sin\theta_P$ at constant $g=1.3$. Bottom: The spectrum
as a function of $u$ at $t=1.6$, and other parameters at the same values.
The spectrum peaks at around $u = 1.1$.}
\label{fig: Wbunch_ut}
\end{figure}

The bunch spectrum has the same dependence on the target size ratio
$g$, the scaled frequency $u$ and the scaled angle of observation 
$t=\gm\sin\theta_P$ as the single particle spectrum. Figure 
\ref{fig: Wbunch_gt} and \ref{fig: Wbunch_ut} show the dependence of
the spectrum on $(g,t)$ and $(u,t)$ respectively. As before, the intensity
initially increases with $g$ but flattens for $g > 2.5$. In the
sequel I will set $g=1.3$ in order to limit the size of the target. Larger
sizes than this if feasible result in significantly larger ODR 
intensities mostly at low frequencies but do not change the intensity
much at high frequencies. Figure 
\ref{fig: Wbunch_ut} shows that the angular spectrum as a function of 
frequency peaks in the vicinity of $u=1$ or $\om=\om_c$ when viewed at
the angle of maximum intensity corresponding to $t=1.6$. 
At other angles, the first peak moves to other values of $u$ and
the peaks are of comparable height. 




\section{Sensitivity to beam parameters}

The beam parameters that we wish to measure with ODR are the beam sizes and
the beam positions. The angular spectral distribution is sensitive to these
parameters and we examine here the dependence on these parameters.
Since the detector will have a finite bandwidth in frequency acceptance,
we include this in our analysis. We define a finite bandwidth spectrum 
by integrating over the frequency as
\beq
\frac{d^2 S}{d\Om d\om} = \int_{\Dl u} du \frac{d^2 W}{d\Om d\om}
\label{eq: Ang_S}
\eeq
Here we assume a 1\% bandwidth for $\Dl u$.

\begin{figure}
\centering
\includegraphics[height=6cm,width=10cm]{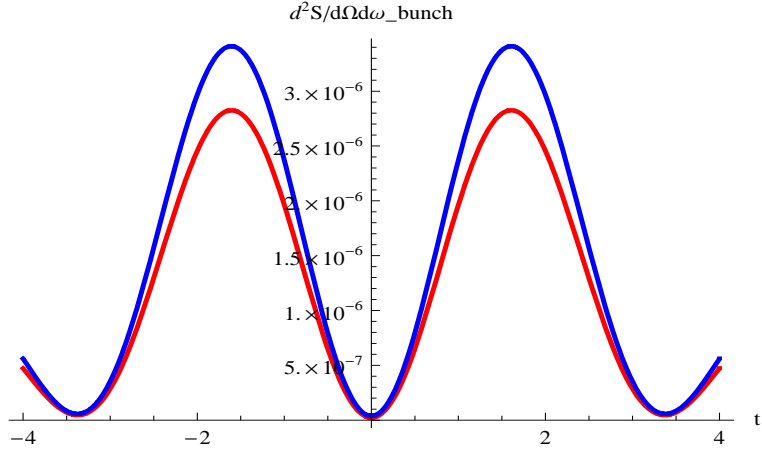}
\caption{The differential angular spectrum of the horizontal
polarization $d^2 S^x/d\Om d\om$ for two different beam sizes. The
red curve is obtained for the scaled beam size $\sg_{x,s}=0.1$ while
the blue curve is obtained for $\sg_{x,s}=0.12$. Both the minimum
and maximum values change with the beam size, the minimum increases
more rapidly on a relative scale. }
\label{fig: Sxbunch_u2_t}
\end{figure}

Consider the sensitivity of the separate polarizations to changes in the beam size.
Figure \ref{fig: Sxbunch_u2_t} shows the horizontal polarization 
intensity $d^2 S^x/d\Om d\om$ 
as a function of the angle variable $t$ for two different values of the scaled
beam size $\sg_{x,s}$. When the beam size is larger filling more of the
aperture, the intensity increases at all observation angles. While
the absolute change in the maximum intensity is larger with larger
beam size, the relative change in the minimum intensity (at $t=0$) is
larger than that in the maximum intensity. 

\begin{figure}
\centering
\includegraphics[height=6cm,width=10cm]{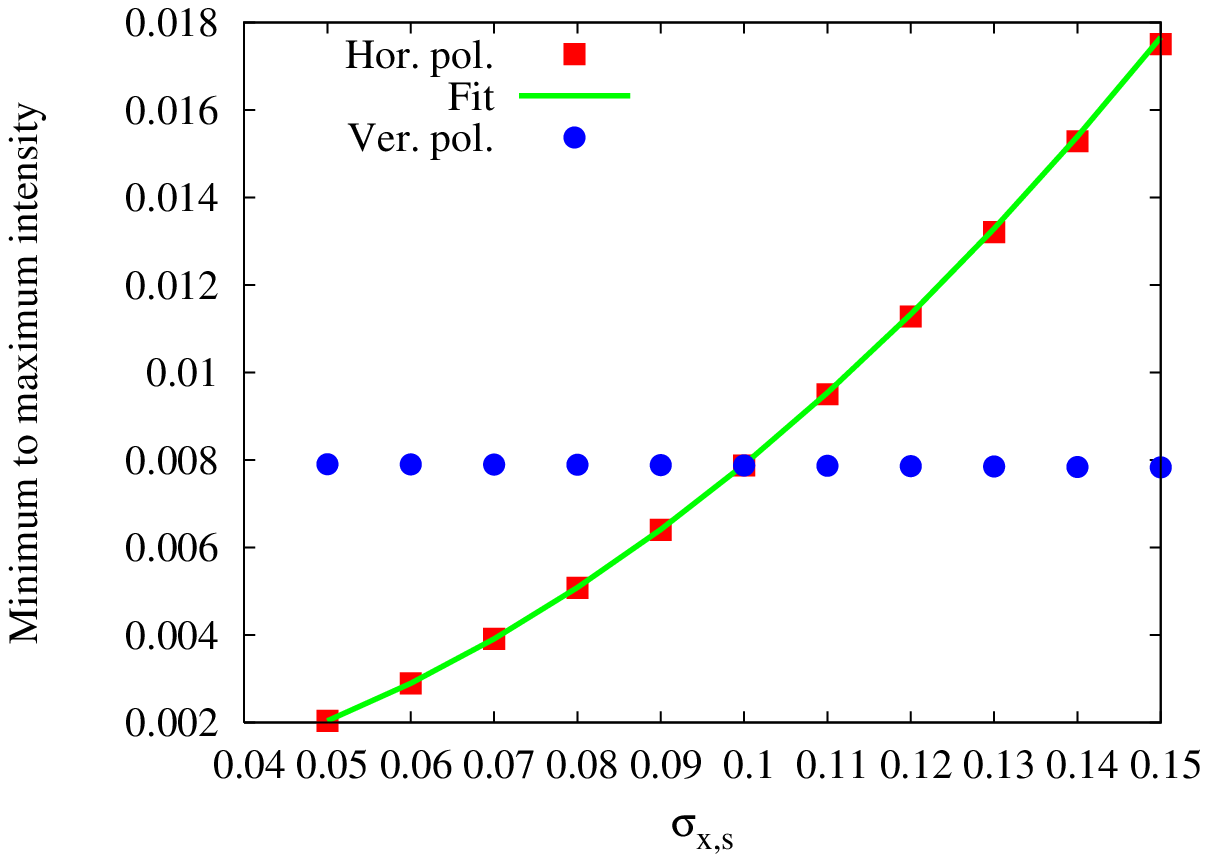}
\includegraphics[height=6cm,width=10cm]{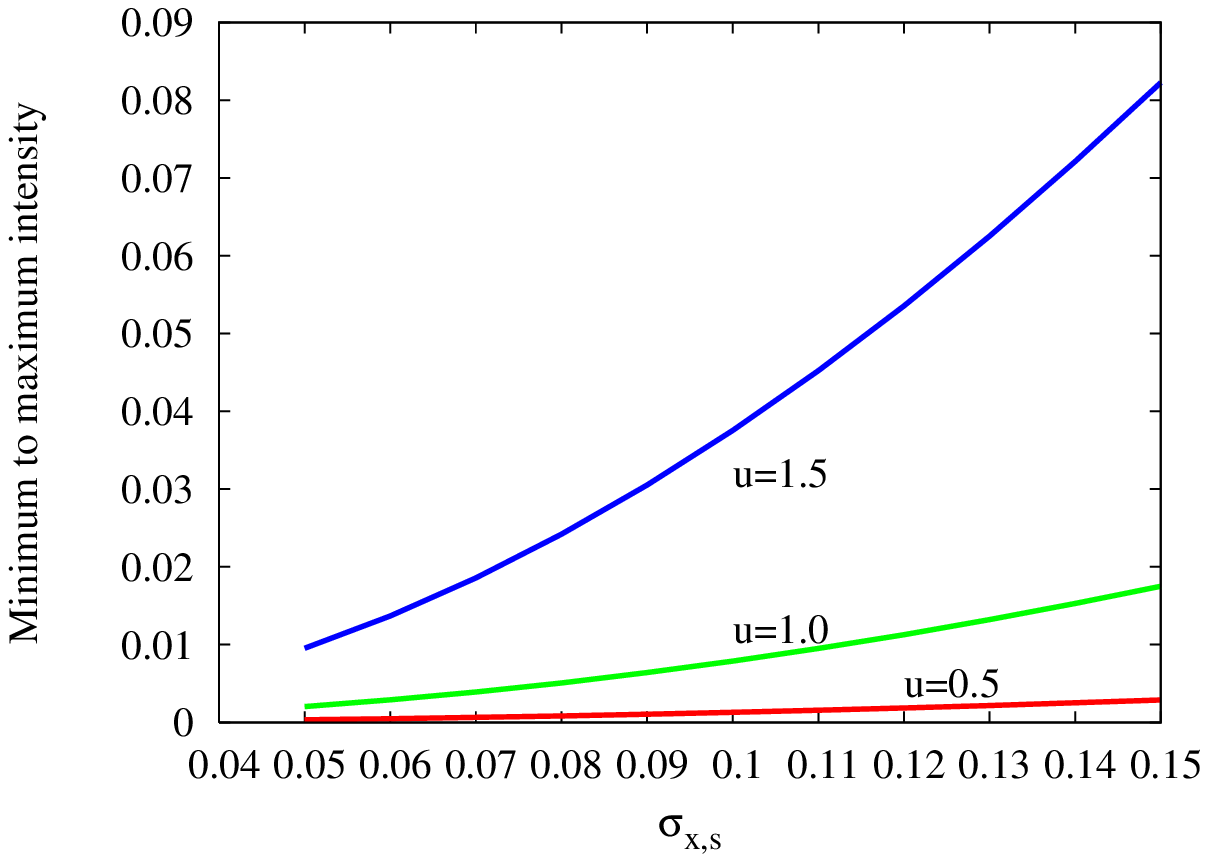}
\caption{Top: The ratio of the minimum/maximum horizontal polarization intensity
and vertical polarization intensity vs the scaled horizontal beam size
$\sg_{x,s}$ at constant $u=1, g=1.3, \sg_{y,s} = 0.1, 
x_{0,s}=0.01=y_{0,s}$. The horizontal polarization intensity increases
quadratically with the beam size as shown by the quadratic fit while the 
vertical polarization intensity
is insensitive to the horizontal beam size changes. Bottom: The minimum to
maximum of the horizontal polarization intensity for two values of $u: 0.5, 1$.
At $u=0.5$, the ratio is smaller which shows that the minimum intensity
is less sensitive to the beam size at the lower frequency. However
the ratio still increases quadratically with the beam size.}
\label{fig: Sx_Sy_bunch_sgx}
\end{figure}

This was exploited in the KEK experiments \cite{KEK_PRL04} where
the beam size was
determined from the ratio of the minimum to the maximum of the angular spectrum
and we apply the same technique here. The top plot in 
Figure \ref{fig: Sx_Sy_bunch_sgx} shows
the ratio of the minimum to maximum of the horizontal and vertical polarization
intensities, $d^2 S^x/d\Om d\om$ and $d^2 S^y/d\Om d\om$ respectively, as a function
of the scaled horizontal beam size $\sg_{x,s}$. 
We mention an important point here: the ratio is independent of the energy, 
bunch intensity etc but depends only on the scaled parameters $(g,u,t)$ 
and scaled 
beam parameters $(\sg_{x,s},\sg_{y,s}, x_{0,s}, y_{0,s})$. Hence as long as the scaled
variables have the same values, these ratios will be the same for the Tevatron, LHC 
etc. The figure shows the data points calculated from the expressions
in Equations (\ref{eq: Ang_Wxy_part}), (\ref{eq: Ang_W_bunch}) and
(\ref{eq: Ang_S}) 
as well as a quadratic fit through these points. The
ratio for the horizontal polarization increases  quadratically with
the horizontal beam size but the vertical polarization is insensitive to 
the change in horizontal beam size. Since the fit to the horizontal 
spectrum is quadratic,  a 1\% increase in the horizontal beam size 
results in a
2\% increase in the minimum to maximum ratio. This suggests that 
{\em if intensity differences
at the level of 2\% can be resolved, then beam size changes at the level
of 1\% can be detected.} The bottom plot in Figure 
\ref{fig: Sx_Sy_bunch_sgx} shows the ratio
for the horizontal polarization at three frequencies: $u = 0.5, 1$. 
The ratio increases quadratically with $\sg_x$ at all frequencies.
Furthermore, the ratio increases with $u$ showing that the sensitivity
of the beam size measurement increases with frequency.

We consider now the sensitivity of the angular spectrum to changes
in the beam offsets. To be useful for diagnostics, the spectrum should
be sensitive to changes in offset which are fractions of a beam size.
We find that when the offsets are in the range $(0-1)\times \sg$, the
maximum of the angular spectrum does not change significantly. However
the minimum of the spectrum at $t=0$ does change rapidly. So we examine
instead the relative change in the minimum intensity as the offset changes.
\begin{figure}
\centering
\includegraphics[height=6cm,width=10cm]{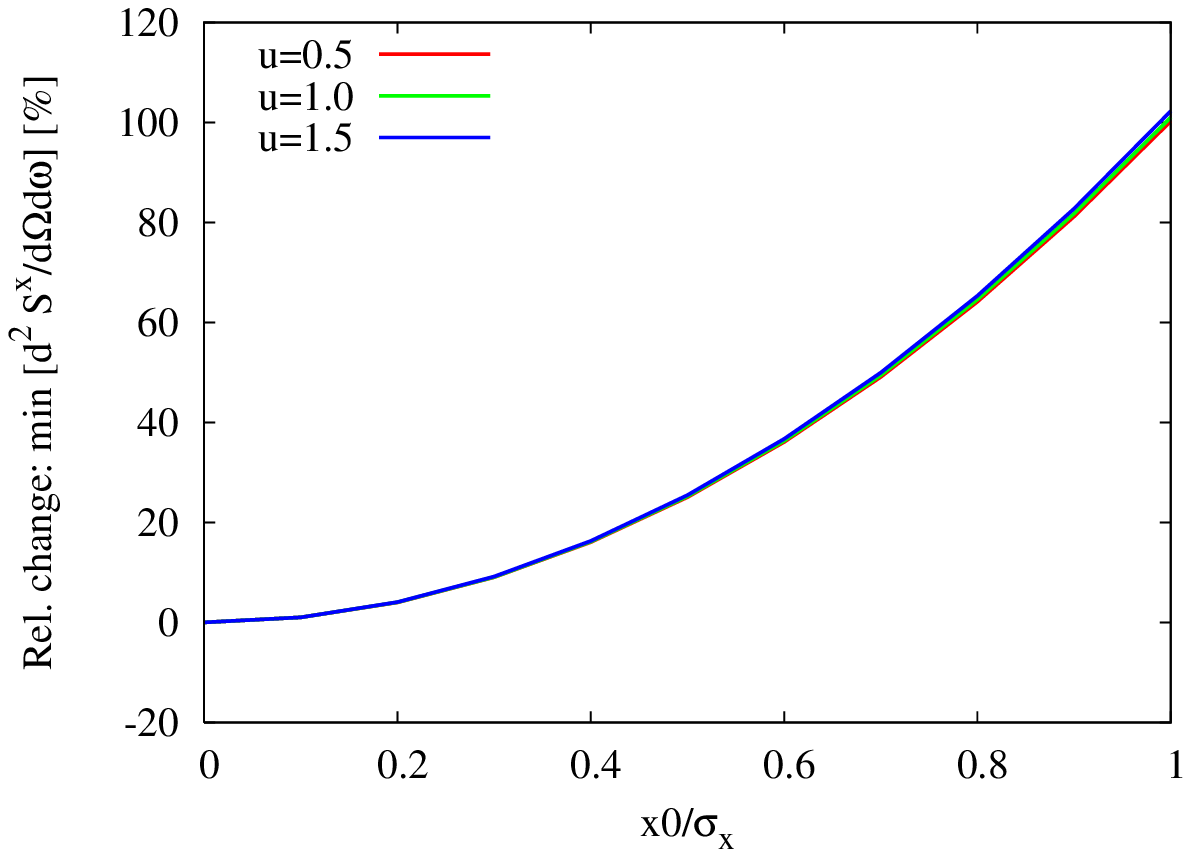}
\includegraphics[height=6cm,width=10cm]{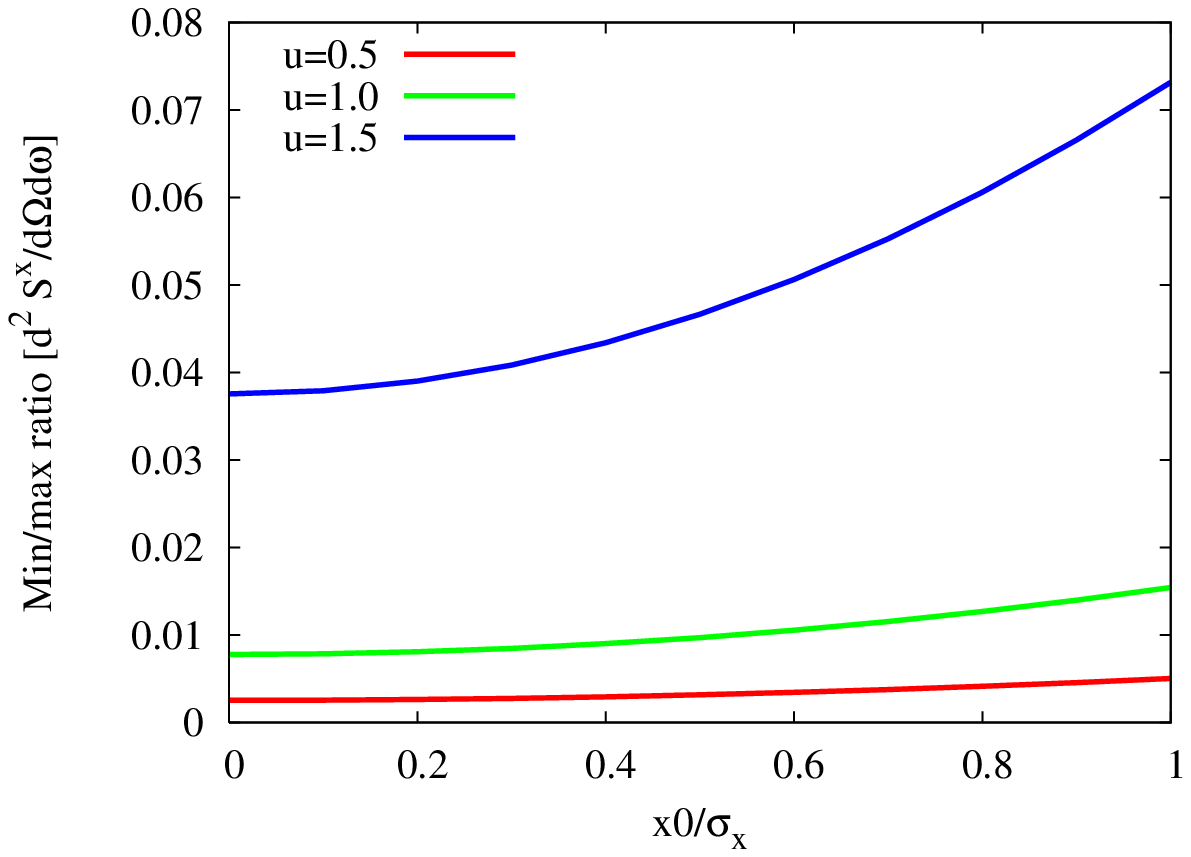}
\caption{Top: The change in the minimum of the horizontal polarization 
intensity in \% as a function of the the horizontal beam offset, shown in 
fractions of the beam size for three values of $u$. This change is not
sensitive to the frequencies in this range. Bottom: Ratio of the minimum 
to the maximum horizontal polarization vs the offset for the same
values of $u$. This ratio increases with the frequencies.}
\label{fig: Sxbunch_x0}
\end{figure}
Figure \ref{fig: Sxbunch_x0} shows the relative change of
$d^2 S^x(t=0)/d\Om d\om$ to changes in the beam offset shown in units
of the beam size for three values of the frequency $u$. 
We observe that as the horizontal offset increases from zero to one times 
the beam size, the minimum horizontal polarization doubles in value.
The relative change in the minimum is nearly the same for the three
values of $u$, so at these frequencies the relative change in the minimum 
is not very sensitive to frequency.
The bottom plot in Figure \ref{fig: Sxbunch_x0} shows the ratio of the
minimum to the maximum of the horizontal polarization. This ratio also
increases quadratically with the offset and more importantly is
sensitive to the frequency,  increasing at higher frequencies.
These results show that if changes in the minimum 
intensity at the level of a few \% can be measured, then changes in
beam offsets of fractions of a beam size can be detected.




\section{Photon yield}

We start by calculating the differential spectrum for a single particle
that is offset from the center of the hole. This will generalize the
results in Section \ref{subsubsec: 0off}. We will then use this result
to calculate the photon yield from a bunch and its dependence on
beam and target parameters.

The differential spectrum for a single particle is found by integrating 
the single particle differential angular spectrum over the solid angle,
\beq
\frac{dW}{d\om}|_{part} = \int \frac{d^2 W}{d\Om d\om}_{part} \sin\theta_Pd\theta_P\;
 d\phi_P = \frac{2}{\gm^2}\int_0^{\gm} dt \int_0^{2\pi} d\phi_P 
 \frac{d^2 W}{d\Om d\om}|_{part} \frac{t}{\sqrt{1 - t^2/\gm^2}}
\eeq
Substituting from Equation (\ref{eq: V_Umn}) and integrating over the
trigonometric functions we find
\beq
\int d\phi_P V_{m,n} = \pi[[I_m'(\frac{u b_s}{\bt})]^2 + (\frac{m\bt}{u b_s})^2
I_m^2(\frac{u b_s}{\bt})](\dl_{m,n} + (-1)^m \dl_{m,-n}) U_{m,m}
\eeq
where $\dl_{m,n}$ is the Kronecker delta and we have used 
$U_{m,-m} = (-1)^m U_{m,m}$. Note that the dependence on $\chi$ has 
disappeared after this integration over $\phi_P$. 
After some simplifications we find
\beq
\frac{dW}{d\om}|_{part}(g,u,b_s) = \frac{q^2}{\pi v} 
\left[ \bar{U}_{00} I_1^2 + 
4\sum_m \bar{U}_{m,m}(I_m^{'2} + (\frac{m\bt}{u b_s})^2 I_m^2) \right]
\eeq
where we have defined the integrated functions
\beq
\bar{U}_{m,m}(g,u,\gm) = \int_0^{\gm} dt 
\frac{t}{\sqrt{1 - t^2/\gm^2}} \frac{1}{(1+t^2)^2} U_{m,m}(g,u,t)
\eeq

The differential spectrum from a bunch is found from
\beq
(\frac{dW}{d\om})|_{bunch}(g,u,\gm) \equiv \int db \; b \int d\chi \rho(b,\chi)
\frac{dW}{d\om}_{part} = \frac{N}{2\pi \sg_{x,s}\sg_{y,s}}
 \int db_s \; b_s \bar{\rho_s}(b_s) \frac{dW}{d\om}_{part}
\eeq
where we have integrated over the scaled density
\beq
\bar{\rho_s}(b_s) = \int d\chi \rho_s(b_s,\chi)
\eeq
From the expression for the density in Equation (\ref{eq: rho_bs_chi}) it
follows that we can write 
\beqr
\rho_s(b_s,\chi) & = & \exp[-\half(\frac{x_{0,s}^2}{\sg_{x,s}^2}
 + \frac{y_{0,s}^2}{\sg_{y,s}^2})]\exp[-b_s^2 \sg_{+,s}]
\exp[-b_s^2 \sg_{-,s}\cos 2\chi]  \nonumber \\
& & \times \exp[\frac{b_s x_{0,s}}{\sg_{x,s}^2}\cos\chi]
\exp[\frac{b_s y_{0,s}}{\sg_{x,s}^2}\sin\chi] 
\eeqr
Here we have defined the scaled beam parameters
\beq
\sg_{\pm,s} = \half[\frac{1}{\sg_{x,s}^2} \pm \frac{1}{\sg_{y,s}^2}]
\eeq
Expanding the exponentials of the trigonometric terms in Bessel functions,
we have
\beqrs
\exp[-b_s^2 \sg_{-,s}\cos 2\chi + 
\frac{b_s x_{0,s}}{\sg_{x,s}^2}\cos\chi +
\frac{b_s y_{0,s}}{\sg_{x,s}^2}\sin\chi] & =  & 
\sum_{j\ge 0}\sum_{k\ge 0} \sum_{l\ge 0}
(-1)^j D_j D_k D_l I_j(b_s^2\sg_{-,s})I_k(\frac{b_s x_{0,s}}{\sg_{x,s}^2})
 \\
& & \times
I_l(\frac{b_s y_{0,s}}{\sg_{y,s}^2})\cos 2j\chi \cos k\chi \cos[l(\pi/2-\chi)]
\eeqrs
where 
\[
D_j  =  1 \;\;\; {\rm for} \;\; j = 0,  \;\;\;\;\;\; 
     D_j =  2 \;\;\; {\rm for} j \ge 1
\]
After integrating over the angle $\chi$, it follows that
\beqr
\bar{\rho}_s(b_s) & = & \frac{\pi}{2}\exp[-\half(\frac{x_{0,s}^2}{\sg_{x,s}^2}
 + \frac{y_{0,s}^2}{\sg_{y,s}^2})] G(b_s) \\
G(b_s) & = & \exp[-b_s^2 \sg_{+,s}] 
\left[ 4 I_0^2(b_s^2\sg_{-,s})I_0(\frac{b_s x_{0,s}}{\sg_{x,s}^2})
 I_0(\frac{b_s y_{0,s}}{\sg_{y,s}^2}) \right. \nonumber \\
& & + \sum_{\stackrel{Not [j=0=l]} j\ge 0}\sum_{l\ge 0}
(-1)^{j+l}D_j D_{2l}I_j(b_s^2\sg_{-,s})I_{2l}(\frac{b_s y_{0,s}}{\sg_{y,s}^2})
\left\{ D_{2|j-l|}I_{2|j-l|}(\frac{b_s x_{0,s}}{\sg_{x,s}^2}) \right. 
\nonumber \\
& & 
\left. \left. + D_{2(j+l)}I_{2(j+l)}(\frac{b_s x_{0,s}}{\sg_{x,s}^2}) \right\} \right]
\eeqr
If the offsets vanish, i.e. $x_0 = 0 = y_0$, this simplifies to
\beq
\lim_{x_0,y_0 \rarw 0} \bar{\rho}_s = 2\pi\exp[-b_s^2 \sg_{+,s}] 
I_0(b_s^2\sg_{-,s})
\eeq

Define the following function obtained after integrating over $b_s$
\beqr
F_{m}(u;\sg_{x,s},\sg_{y,s},x_{0,s},y_{0,s}) & = & \int_0^1 db_s \; b_s G(b_s)
[ (I_m'(\frac{u b_s}{\bt}))^2 + (\frac{m\bt}{u b_s})^2
I_m^2(\frac{u b_s}{\bt})] 
\eeqr
In the limit that the offsets vanish, this simplifies to 
\beqr
\lim_{x_0,y_0 \rarw 0} F_{m} & = & 4
\int_0^1 db_s \; b_s \exp[-b_s^2 \sg_{+,s}] 
I_0^2(b_s^2\sg_{-,s}) [ (I_m'(\frac{u b_s}{\bt}))^2 + (\frac{m\bt}{u b_s})^2
I_m^2(\frac{u b_s}{\bt})] 
\eeqr

The differential spectrum from a bunch is obtained from
\beq
(\frac{dW}{d\om})|_{bunch} = \frac{q^2}{2\pi v} \frac{N}{\sg_{x,s}\sg_{y,s}}
\exp[-\half(\frac{x_{0,s}^2}{\sg_{x,s}^2} + \frac{y_{0,s}^2}{\sg_{y,s}^2})]
\left\{ \bar{U}_{0,0}F_0 + 4\sum_{m\ge 1} \bar{U}_{m,m} F_{m} \right\}
\eeq

The photon yield into a bandwidth $\Dl \om$ is 
\beq
\fbox{\rule[-.5cm]{0cm}{1cm} 
$\displaystyle 
\Dl N_{ph} = \frac{\alp_f}{2\pi\bt} \frac{N}{\sg_{x,s}\sg_{y,s}}
\exp[-\half(\frac{x_{0,s}^2}{\sg_{x,s}^2} + \frac{y_{0,s}^2}{\sg_{y,s}^2})]
\left\{ \bar{U}_{0,0}F_0 + 4\sum_{m\ge 1} \bar{U}_{m,m} F_{m} \right\}
\frac{\Dl \om}{\om} $}
\eeq

This is the photon yield from a single bunch per turn over the full 
$4\pi$ solid angle, the number of
photons intercepted by the detector will be reduced by the acceptance of
the detector.

\begin{figure}
\centering
\includegraphics[height=6cm,width=10cm]{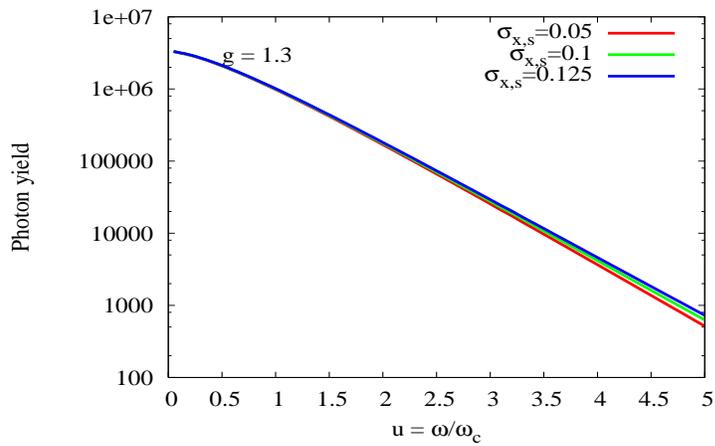}
\caption{Photon yield (on a log scale) from a single bunch per turn into a 
1\% bandwidth as a function of the
scaled frequency $u$ for three values of the scaled beam size $\sg_{x,s}$.
In all cases $\sg_{y,s}= \sg_{x,s}$. Tevatron bunch intensity and energy
were used in this calculation. The photon yield is not very sensitive
to the beam size at low frequencies but the relative difference in yield 
increases at higher frequencies.}
\label{fig: photoncount}
\end{figure}
Figure \ref{fig: photoncount} shows the photon yield (plotted on a 
logarithmic scale) from a single bunch
per turn into a 1\% bandwidth as a function of the scaled frequency 
$u = \om/\om_c$ for three 
values of the scaled beam size $\sg_{x,s}$. We have set the offsets 
$x_0, y_0$ to zero for this calculation. The photon 
count is about $10^6$ photons per bunch per turn at 
$u=1$ which should yield a detectable signal. 
The dependence of the photon yield on frequency can be fit to
exponential curves with two different exponents below and above $u =1$.
For example, for $\sg_{x,s} = 0.125$ we find that
\beqr
 \Dl N_{ph} & \sim & \exp[- 1.28 u ] , \;\;\;\;\; u < 1 \nonumber \\
              & \sim & \exp[- 1.82 u ] , \;\;\;\;\; u \ge 1 
\eeqr
These exponents are not very sensitive to the beam size $\sg_{x,s}$, 
e.g the exponents below and above $u=1$ are $(-1.29u, -1.85u)$ for 
$\sg_{x,s}=0.1$. 

 The photon yield does not depend sensitively on the beam size at low
frequencies but at higher frequencies, the sensitivity to beam size
increases.

The scaled frequency $u$ can be converted to a physical frequency by
making a choice of the inner radius $a_{in}$ of the hole. If we assume
$\sg_{x,s} = 0.125 = \sg_{y,s}$ or 
$a_{in} = 8\sg$ and $\sg \approx 0.4$mm at the C0 location in the Tevatron,
then the critical wavelength $\lm_c$ corresponding to the critical
frequency $\om_c$ is $\lm_c = 2\pi a_{in}/\gm = 19\mu$m. This is in
the infra-red range. Detection in the optical range seems to be ruled out
since we find that at $u=19$ or $\lm = 1\mu$m, the photon count is
about 10$^{-8}$/bunch/turn or practically zero. 
We can increase the photon yield by increasing the
target size to say $g=2.5$, the optimal value found earlier.
\begin{figure}
\centering
\includegraphics[height=6cm,width=10cm]{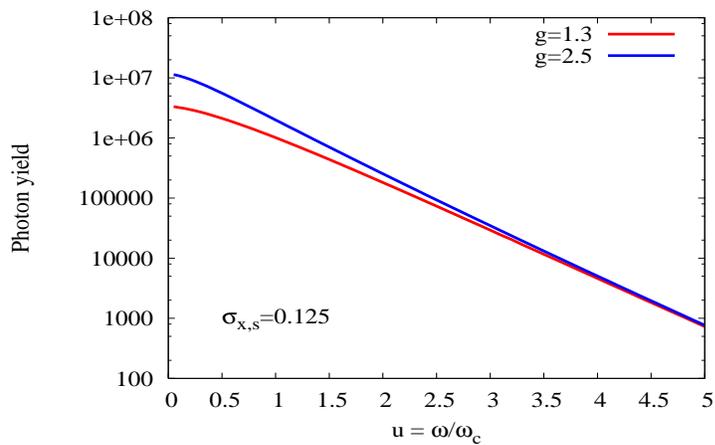}
\caption{Photon yield (on a log scale) from a single bunch per turn into a 
1\% bandwidth as a function of the
scaled frequency $u$ for two values of the target size $g$.
Here the scaled beam size is chosen as $\sg_{x,s}= 0.125 = \sg_{y,s}$. }
\label{fig: photoncount_g2.5}
\end{figure}
Figure \ref{fig: photoncount_g2.5} shows the yields for $g=1.3, 2.5$.
At low frequencies $u < 1.5$ the difference is significant but not so
at higher frequencies. There is therefore no advantage to be gained
with a larger target size if we choose to operate at frequencies above
$u > 2$. 

We can use the photon yield to choose the frequency at which to observe
the ODR. In practice the minimum photon yield will be determined by the 
efficiency and resolution of the camera and the level of the background
synchrotron radiation which should be below the ODR photon yield.
If for example we set the minimum photon yield to be $10^4$ 
photons/bunch/turn into a 1\% bandwidth, then the highest scaled frequency
from the plots above is $u = 3.5$. For the Tevatron, this implies that
observable wavelength has to be at or above 5.4$\mu$m. If we integrate the
signal over all 36 bunches and the bandwidth is greater than 1\%, then
the wavelength could be reduced some more. However it will be in the
few microns range and not in the optical range.

\section{OTR vs ODR spectrum}

OTR is generated when the beam goes through the material of the target.
The expressions for the OTR spectrum can be found from the ODR spectrum
by taking the limit $a_{in} \rarw 0$. With OTR we cannot define a
critical frequency $\om_c$ and OTR observations show that there is no frequency
at which the frequency peaks. This is a qualitative difference from the
ODR spectrum. Similarly we cannot also define universal expressions for
the OTR spectrum which depend only on dimensionless parameters.

In this section we will briefly discuss the OTR single particle spectrum
as a limiting case of the ODR spectrum. 
First consider the single particle at the center of the OTR target. 
From Equation (\ref{eq: angspect_1part_0off}) it follows that the differential angular spectrum in the far-field is given by
\beqr
\frac{d^2 W^{OTR}}{d\Om d\om}|_{part} & = & \half \bt c (\frac{k q \alp}{\pi v})^2 
\frac{1}{[\bar{k}^2+\alp^2]^2} T(a; \bar{k})^2 \nonumber \\
& = & \half \frac{q^2 \gm^2}{\pi^2 v} \frac{a^2}{(1+t^2)^2}
[\bar{k}J_0(\bar{k} a)K_1(\alp a) + \alp J_1(\bar{k} a)K_0(\alp a)]^2
\eeqr
Here $a$ is the radius of the target, $\bar{k} = k\sin\theta_P$, $\alp = k/\gm$.

\begin{figure}
\centering
\includegraphics[height=6cm,width=10cm]{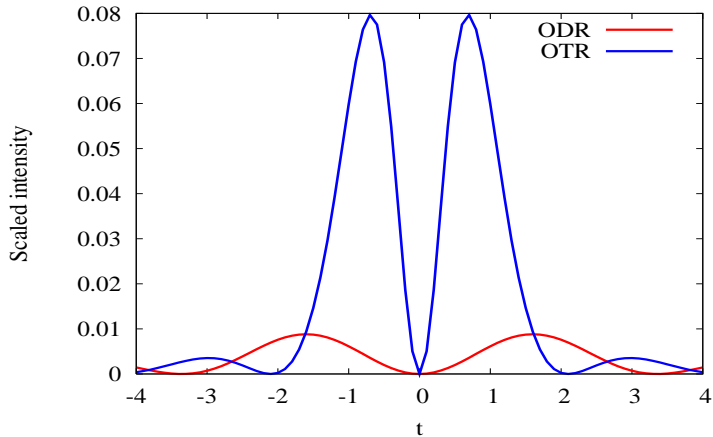}
\caption{Comparison of ODR and OTR intensities for a single particle
centered on the target. The outer radius $a_{out}$ is assumed to be the
same in both cases. }
\label{fig: ODR_OTR}
\end{figure}
Figure \ref{fig: ODR_OTR} shows the dependence of the single particle
OTR and ODR spectrum on the angular variable $t$. For the parameters
chosen here ($a_{out} = 1.3 a_{in}$, $u = 1$), the maximum OTR
intensity is about 8 times larger than the ODR intensity.
Of this,  a factor of 2.45 is due to the different areas of the material in 
the target for OTR and ODR. Furthermore,
the OTR spectrum peaks at a smaller angle $t \approx 1 \Rarw 
\theta_P \approx 1/\gm$ while the ODR spectrum peaks at $t \approx 1.6$.
This figure shows that ODR is beamed in a broader cone and at a larger
angle than OTR. 
For the single particle spectrum with the particle offset from the center,
Equation (\ref{eq: spectrum_1part}) is still applicable for the differential
angular spectrum except that $U_{mn}$ defined in Equation (\ref{eq: V_Umn}) 
simplifies to 
\beqr
U_{m,n} & = & a^2
[\bar{k}J_{m-1}(\bar{k} a)K_m(\alp a) + \alp J_m(\bar{k} a)K_{m-1}(\alp a)] 
\nonumber \\
& & \times [\bar{k}J_{n-1}(\bar{k} a)K_n(\alp a) + \alp J_n(\bar{k} a)K_{n-1}(\alp a)] 
\eeqr

The OTR bunch spectrum can be similarly found by taking the limit $a_{in}\rarw 0$ 
in the appropriate expressions above.

\section{Experimental Issues}

The devices and experimental conditions needed to observe ODR in the Tevatron
requires a separate detailed discussion. Here we will only mention some
of the issues. Synchrotron radiation from the upstream dipoles hitting the
target is an important source of background and needs to be mitigated.
Preliminary calculations \cite{Randy} show that the level of background
at a target near the C0 point in the Tevatron is less than the anticipated
ODR flux. A mask placed upstream of the target may help to reduce this
background to acceptable levels. The wavelength at which to observe the
ODR depends on several competing factors. At longer wavelengths the 
ODR flux is higher but far infra-red detection is less sensitive, slower and 
complicated by other matters such as choice of windows
which are sufficiently transparent at longer wavelengths. The synchrotron
radiation background also increases at longer wavelengths. 
A satisfactory compromise might be in the vicinity of 5$\mu$m. 
Given the likely speed with which the ODR images will be acquired, it
is unlikely that bunch by bunch and turn by turn imaging will be possible
with the ODR monitor. Averaging over turns will most likely be necessary
but it should be possible to update images on the time scale of seconds.
Calibration of the ODR
measurements requires measurements by other beam imaging devices nearby.
At C0, the synchrotron light monitor is relatively close and would be
suitable for calibration of the ODR monitor. If a circular target is used,
it would likely be made of two semi-circular halves which will be moved in
towards the beam on separate stepper motors. These two halves will need to
be precisely aligned and their positions measured with respect to the beam.
It is very likely that due to the nature of the helical orbits, the target 
in the Tevatron will only be suitable for one beam. We would choose the 
proton beam for ease of availability and greater intensity. It would be
preferable but not essential to choose the longitudinal location of the 
target so that both beams are not present simultaneously to avoid any
effects from their parasitic interaction. Independently of this, the
circular target may not be suitable if the helical orbits are separated by 
several beam sizes at the target. In this case the desired beam may have
the required separation from the target but the opposing beam will not
be far enough from the target. This could be avoided by having the
two halves of the target separated by a gap. Instead of semi-circular
foils, rectangular foils on either side may be preferable in this case.

 We envisage that if initial measurements are successful, the ODR monitor
could be used a passive device monitoring beam parameters and their changes
during the length of a store. Here other operational challenges will arise.
For example, beam motion which changes the position relative to the target
will need to be included in the automated ODR measurement. Wake fields due
to the target and heating of the target by beam induced fields are likely
to be negligible but need to be considered. Some of these same
issues have arisen and been resolved with the use of the pick-off mirrors 
for the synchrotron light mirror and will benefit from that experience. 
A detailed account of these and other relevant issues will appear elsewhere.

\section{Conclusions}

Our main concern here was the far field ODR spectrum from a round hole
in a collider and specifically the Tevatron. We list the main conclusions
\bit
\item Existence of a critical frequency and universal curves.

There is a critical frequency $\om_c$ associated with ODR at which
the angular spectrum intensity peaks. There is no such frequency with OTR
This is a well known phenomena. However we have also shown, something
not previously recognized, that
the spectrum for a round hole depends only on dimensionless parameters.
Hence the results seen in Figures \ref{fig: Wbunch_gt} to 
\ref{fig: Sxbunch_x0}, especially the sensitivities, are universally 
applicable to all machines
when these dimensionless parameters have the same values. 

\item Sensitivity to beam sizes.

 The ratio of the minimum to maximum intensity is very sensitive to the 
beam size as seen in Figure \ref{fig: Sx_Sy_bunch_sgx}. 
The sensitivity increases with observation frequency.
The horizontal polarization is sensitive to the horizontal beam size
and similarly for the vertical plane. This differs from the dependence
with rectangular slits and straight edges.

\item Sensitivity to beam offsets

We found that the minimum of the angular spectrum along the direction of
specular reflection is most sensitive to changes in the beam offset.
The ratio of the minimum to the maximum can also be used, this has the
added advantage of higher sensitivity at increasing frequencies.
Again, we found very good sensitivity (seen in Figure 
\ref{fig: Sxbunch_x0}) to changes in beam offset of fractions
of a beam size.

\item Photon yield.

We calculated the photon yield from a single Tevatron bunch at 980 GeV.
The photon yield decreases exponentially fast with frequency. Assuming 
an 8$\sg$
separation between the beam and the target, the critical wavelength is
19$\mu$m. The calculation predicts that photon yields of $\approx 10^4$ 
photons/bunch/turn into a 1\% bandwidth will be obtained at about 
a 5$\mu$m wavelength. Detection at optical frequencies does not seem
feasible. It is clear that the lower wavelength limit for 
observable ODR signals in the far-field
is at a few microns, in the infra-red regime.

\eit

In this paper we did not consider the near-field spectrum in much detail
except briefly in Section \ref{subsubsec: off}.
That discussion as applicable to the spectrum from a bunch and a detailed 
discussion of experimental details will appear separately

\vspace{2em}

\noi {\bf \Large Acknowledgments}

I thank Alex Lumpkin, Vic Scarpine, Randy Thurman-Keup and Manfred Wendt
for several useful discussions and for sharing their results with me.

\end{document}